\documentclass[sigchi, authorversion]{acmart}

\usepackage{comment}
\usepackage{enumitem}
\usepackage[english]{babel}
\usepackage{blindtext}
\usepackage{booktabs}
\usepackage{svg}
\usepackage{multirow}
\usepackage{graphicx}
\usepackage{gensymb}
\usepackage{longtable}
\usepackage{tabularx}
\usepackage{makecell}
\usepackage{arydshln} 

\AtBeginDocument{%
  \providecommand\BibTeX{{%
    \normalfont B\kern-0.5em{\scshape i\kern-0.25em b}\kern-0.8em\TeX}}}

\copyrightyear{2026}
\acmYear{2026}
\setcopyright{cc}
\setcctype{by}
\acmConference[CHI '26]{Proceedings of the 2026 CHI Conference on Human Factors in Computing Systems}{April 13--17, 2026}{Barcelona, Spain}
\acmBooktitle{Proceedings of the 2026 CHI Conference on Human Factors in Computing Systems (CHI '26), April 13--17, 2026, Barcelona, Spain}
\acmPrice{}
\acmDOI{10.1145/3772318.3790401}
\acmISBN{979-8-4007-2278-3/2026/04}

\definecolor{RevisionColor}{rgb}{0,0,0}
 \newcommand{\revised}[1]{\textcolor{RevisionColor}{#1}}

\begin{document}
\acmSubmissionID{3620}

\title[Running into Traffic]{Running into Traffic: Investigating External Human-Machine Interfaces for Automated Vehicle-Runner Interaction}

\author{Ammar Al-Taie}
\email{ammar@kaist.ac.kr}
\orcid{0000-0002-5156-6245}
\affiliation{%
  \institution{Information and Electronics Research Institute, KAIST}
  \country{Republic of Korea}
}

\author{Thomas Goodge}
\email{thomas.goodge@glasgow.ac.uk}
\orcid{0000-0001-6229-7936}
\affiliation{%
  \institution{Glasgow Interactive Systems Section, School of Computing Science, University of Glasgow}
  \country{United Kingdom}
}

\author{Shaun Macdonald}
\email{shaun.macdonald@glasgow.ac.uk}
\orcid{0000-0001-6519-318X}
\affiliation{%
  \institution{Glasgow Interactive Systems Section, School of Computing Science, University of Glasgow}
  \country{United Kingdom}
}

\author{Ian Oakley}
\email{ian.r.oakley@gmail.com}
\orcid{0000-0001-5834-8577}
\affiliation{%
  \institution{School of Electrical Engineering, KAIST}
  \country{Republic of Korea}
}

\author{Stephen Brewster}
\email{stephen.brewster@glasgow.ac.uk}
\orcid{0000-0001-9720-3899}
\affiliation{%
  \institution{Glasgow Interactive Systems Section, School of Computing Science, University of Glasgow}
  \country{United Kingdom}
}

\begin{abstract}
Automated vehicles (AVs) must communicate their yielding intentions to pedestrians at crossings. External Human-Machine Interfaces (eHMIs, on-vehicle displays) are promising solutions, but were primarily tested with walking pedestrians. Runners are a significant pedestrian group who move faster and face distinct bodily and perceptual demands, raising questions about how pedestrian activity influences eHMI use. We conducted an outdoor study using an augmented reality simulator. Participants navigated a virtual crossing while walking and running; an approaching AV displayed one of three eHMIs: red/green colour-changing lights, animated cyan lights, or no-eHMI. No-eHMI consistently underperformed. Walkers mostly stopped and validated eHMI signals with vehicle behaviour; they processed both eHMI animations and colour changes effectively. Runners experienced greater time pressure to cross, increasing reliance on the eHMI over vehicle behaviour. They preferred colour changes over animation for rapid decisions. These findings are crucial for promoting eHMI inclusivity and physical wellbeing as AVs join our roads.
\end{abstract}

\renewcommand{\shortauthors}{Al-Taie et al.}



\begin{teaserfigure}
    \centering
  \includegraphics[width=0.75\linewidth]{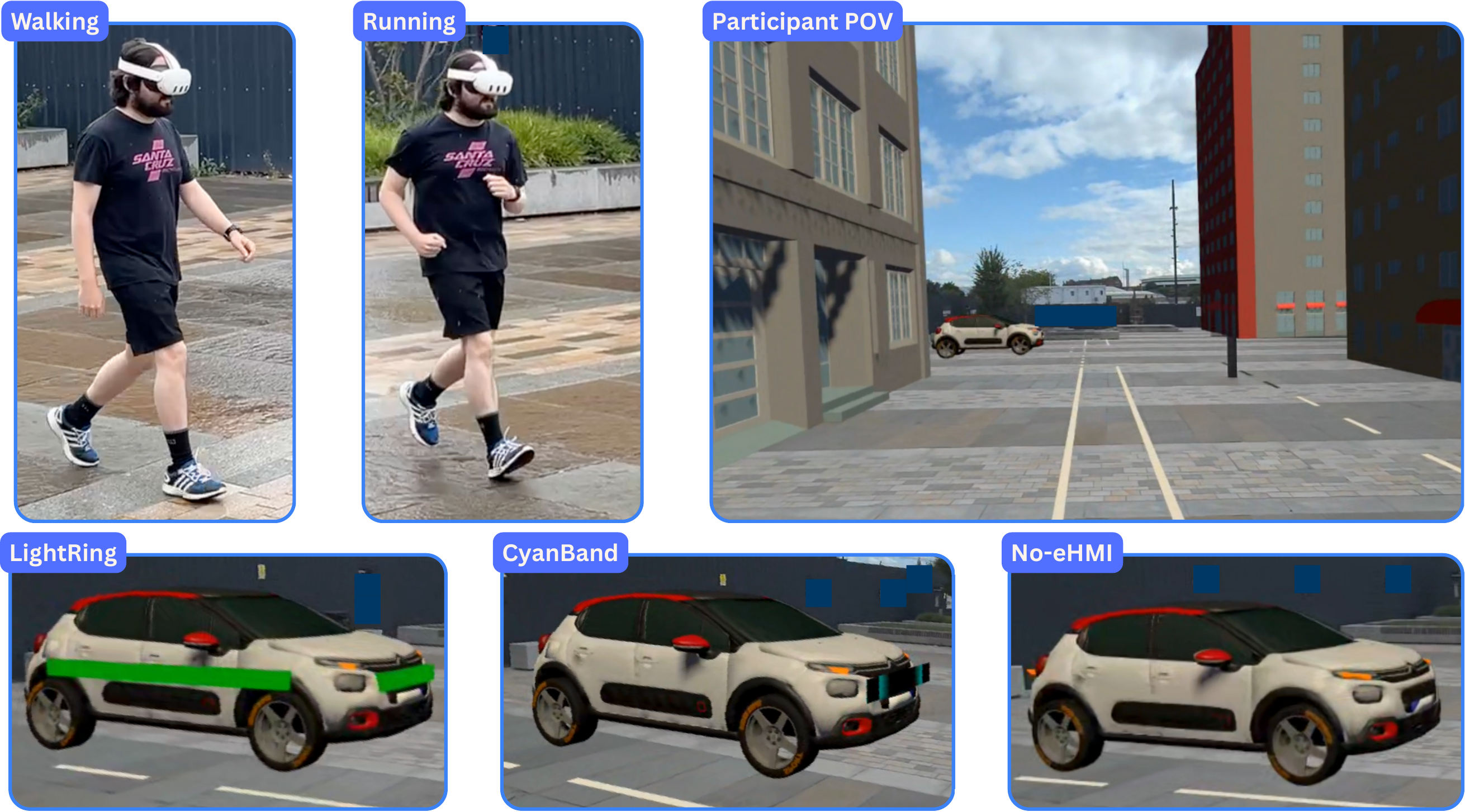}
  \caption{Our study setup. The top images show a participant using an AR simulator to navigate a virtual crossing while either walking or running, and their point-of-view of an augmented urban environment in the simulator. The bottom images show the tested eHMI conditions: LightRing (red/green colour changing lights), CyanBand (animated cyan lights on the front) or no eHMI at all.}
  \label{fig:teaser}
\end{teaserfigure}

\maketitle

\section{Introduction}

The pedestrian road user group includes any person travelling on foot \cite{UK2024TheRules}, such as walkers or runners. They are vulnerable road users (VRUs) who commonly encounter vehicles when crossing roads \cite{Hollander2021AReview}. To cross safely, pedestrians must understand whether a driver intends to yield or continue driving \cite{Haddington2014InteractionTraffic}. Relying on the vehicle's driving and braking behaviours could be ambiguous or misleading \cite{Markkula2020DefiningTraffic}. Instead, pedestrians often depend on social interaction with drivers to confirm intent. For example, a driver may use a hand gesture to signal a pedestrian to proceed \cite{Dey2017TheVehicles}, or a pedestrian may seek eye contact to confirm they have been seen \cite{Dey2019GazeStudies}. \revised{Such interactions are crucial for pedestrians to make informed crossing decisions and avoid collisions \cite{DepartmentforTransport2024ReportedFactsheet}. However, automated vehicles (AVs) are now being deployed in urban traffic, with over 1 million trips completed each month in the USA alone in 2025 \cite{Waymo2025ScalingManufacturing}.} \revised{AVs remove the need for human drivers, but also diminish important social cues from behind the wheel \cite{Rasouli2020AutonomousPractice}. This shifts road interaction from interpersonal to human-machine \cite{Krizsik2022SocialVehicles}, prompting a growing body of Human-Computer Interaction (HCI) research to investigate clear, explicit communication between AVs and pedestrians \cite{Tran2021AInteraction}. The current consensus supports the use of external Human-Machine Interfaces (eHMIs): displays mounted on the vehicle's exterior to communicate with other road users \cite{Dey2020TamingInterfaces}, such as lights on the bumper \cite{Dey2020ColorPedestrians} or speakers on the roof \cite{Dey2024Multi-ModalInteraction}. Placing eHMIs on vehicles enables pedestrians to receive AV signals from a familiar location \cite{Al-Taie2023KeepTraffic}, without requiring them to carry any additional devices \cite{Berge2022DoStudy}. However, these eHMIs become shared public displays that must be inclusive of diverse road user types and behaviours \cite{Hollander2021AReview}.}

\revised{Consequently, research has investigated eHMI requirements for cyclists \cite{Hou2020AutonomousPromise}, e-scooter riders \cite{Matviienko2022E-ScootAR:Reality}, drivers \cite{Rettenmaier2020AfterScenarios} and other road user types \cite{Hollander2021AReview}. However, pedestrian research remains focused on those moving at walking speeds \cite{Dey2020TamingInterfaces}, overlooking a significant subgroup: runners. This omission is problematic because interactions at crossings are not only perceptual but embodied: movement, urgency, and attentional load could shape how signals are interpreted and acted upon \cite{Kirsh2013EmbodiedDesign, Marentakis2024SituationalInterfaces}. Running is the most popular physical activity worldwide \cite{vanRheden2024GestureShirt:Running}, with over 621 million regular participants \cite{RunningWithGrit2025StrikingLife}, and participation continues to rise as international marathons consistently report record entry numbers each year \cite{LondonMarathonEvents2025NewMarathon, Si-Nae2025KoreasMeasures}. Many runners train in urban environments and frequently navigate road crossings with approaching vehicles \cite{Deelen2019AttractiveEnvironment.}. However, most HCI work on runners has focused on sports and fitness rather than road safety \cite{Mencarini2019DesigningInteraction, Jensen2014RunningTechnology, Kuru2016ExploringTechnology, Muijlwijk2024BenefitsRunning}, so the eHMI requirements to support runners are unknown. While walkers and runners travel on foot \cite{UK2024TheRules} and encounter AVs in similar crossing scenarios \cite{Berge2024TriangulatingData}, runners move faster than walkers \cite{Yao2022RecognizingSpeeds}, reducing the time available to interpret eHMI signals \cite{Seuter2017RunningTechnology}. HCI theories on situational impairment \cite{Marentakis2024SituationalInterfaces} suggest that a user's speed limits how they engage with interfaces designed for slower contexts \cite{Sarsenbayeva2017ChallengesDevices}, raising doubts about the generalisability of walker-centric eHMIs for runners \cite{Hollander2021AReview}. Runners also exhibit distinct movement patterns from walkers: they take longer strides \cite{Alexander1996WalkingRunning}, spend less time in ground contact \cite{Yao2022RecognizingSpeeds}, and display greater vertical body motion \cite{Jin2024LowerTransitions}. Theories of embodied cognition \cite{Kirsh2013EmbodiedDesign} indicate that these bodily behaviours may determine how runners interact with AVs and interpret eHMIs \cite{Bergstrom-Lehtovirta2011TheInterface}. This raises crucial questions about how the speed and embodied behaviours of pedestrians influence eHMI use, and how human–machine communication can support rapid, safety-critical decisions made while running.}

To address this, we conducted a user study where participants used an augmented reality (AR) simulator to navigate a virtual crossing while walking or running in the real world. A virtual AV approached the crossing and either yielded or did not yield. It communicated its intentions using one of three eHMIs: LightRing \cite{Al-Taie2024LightInterfaces} (red/green colour changing-lights around the vehicle), CyanBand \cite{Dey2020ColorPedestrians} (sweeping cyan animations on the vehicle's bumper), or No-eHMI at all. \revised{We found that crossing behaviours and eHMI use were shaped by the pedestrian's motivation to cross and the embodied constraints of their physical activity. Participants described walking as a commuting or leisurely activity, where slowing down or stopping requires little effort. This gave them sufficient time to interpret and verify eHMI signals against the AV’s driving behaviour. Consequently, walkers reported greater trust and confidence in AV intentions and consistently stronger feelings of safety, even in situations where the AV did not yield. This added time also allowed walkers to use both animation-based and colour-changing eHMIs effectively. In contrast, running was framed as an exercise activity with specific performance goals. Participants found it physically costly to slow down and preferred to maintain their pace, resulting in fast and mobile AV interactions. Here, runners relied heavily on the eHMI to make rapid crossing decisions, often without considering the AV’s implicit driving cues. This resulted in higher perceived risk, lower perceived safety, and more collision-prone behaviours. Runners preferred colour-changing eHMIs, as animation-based displays demanded more processing time than their running pace allowed.} Based on these insights, we discuss how interfaces can be inclusive of diverse pedestrian movement behaviours, which is essential for the large-scale adoption of AVs \cite{Anderson2016AutonomousPolicymakers} and safe active mobility in urban environments \cite{UnitedNations2025SustainableGoals}. This paper contributes:

\begin{itemize}
    \item The first user study to consider runners and AVs, establishing runners as a distinct user group from walkers;
    \item Empirical evidence showing how the distinct movements of walkers and runners influence eHMI use across AV-yielding behaviours;
    \item Design guidelines for inclusive eHMIs that accommodate pedestrians exhibiting distinct movement behaviours;
    \item A novel AV interface design based on the guidelines, demonstrating how AV communication can adapt to diverse pedestrian movement behaviours.
\end{itemize}
\section{Related Work}

The AV interaction problem is widely documented in prior work. The road is a ``shared social space'' \cite{Latham2019AutonomousTechnology}, with road users exchanging social cues to negotiate right of way \cite{Markkula2020DefiningTraffic, Haddington2014InteractionTraffic, Dey2017TheVehicles}. However, AVs will replace human drivers and remove these social interactions, posing new safety risks \cite{Rasouli2020AutonomousPractice}. \citeauthor{Markkula2020DefiningTraffic}~\cite{Markkula2020DefiningTraffic} contributed a taxonomy of the cues that AVs must recognise and communicate. They divided the cues into \textit{implicit cues}, which include the vehicle's driving and breaking behaviours, and \textit{explicit} ones, which are direct signals explaining a vehicle's intentions. AVs can utilise implicit cues by adjusting their speed during space-sharing conflicts, but explicit cues must be compensated for clear communication \cite{Rasouli2020AutonomousPractice, Dey2020TamingInterfaces, Lee2021RoadVehicles}. \revised{This was confirmed through real-world observations; \citeauthor{Brown2023TheTraffic}~\cite{Brown2023TheTraffic} analysed videos of AV-walker interactions, and found many issues and ambiguities due to AVs being unable to communicate their intentions beyond braking or accelerating. In one representative example, an AV slowed down to yield, but the walker mistrusted its intentions and waved to signal the AV to pass. However, the AV exhibited hesitant behaviours of repeatedly braking and accelerating before eventually proceeding. Such behaviours are uncommon in regular traffic \cite{Kauffmann2018WhatScenario, Haddington2014InteractionTraffic}, showing that AVs must compensate for driver social cues to be accepted.}

\subsection{eHMIs are Promising Solutions}

\revised{Early work investigated the form factors and placements AV interfaces should adopt to compensate for social cues. \citeauthor{Colley2020AVehicles}~\cite{Colley2020AVehicles} contributed an AV interface design space based on a literature review and focus group with experts. They identified the pedestrian, road infrastructure, and vehicle, as potential interface placements. On-vehicle interfaces (eHMIs) would be particularly advantageous because this allows road users to receive AV intentions without carrying any additional devices \cite{Berge2022DoStudy, Al-Taie2022TourBehaviour}.} \citeauthor{Dey2019GazeStudies}~\cite{Dey2019GazeStudies} conducted an eye-tracking study in which participants stood at a crossing and a vehicle with a hidden human driver approached. They analysed where participants looked, and found that their gaze transitioned from the vehicle bumper (for implicit cues) to the windscreen, where they typically receive social cues, as the vehicle moved closer. This suggests that the vehicle itself offers a familiar placement for interfaces, motivating the use of on-vehicle eHMIs. \citeauthor{Al-Taie2023KeepTraffic}~\cite{Al-Taie2023KeepTraffic} conducted a similar study with cyclists who wore eye-tracking glasses while riding in real traffic. Cyclists consistently looked at vehicle bodies and windscreens. This indicates that eHMIs are potentially inclusive solutions that match the natural behaviours of different road users, not just walkers. 

To begin developing eHMIs, \citeauthor{Dey2018InterfaceUsers}~\cite{Dey2018InterfaceUsers} conducted design sessions with HCI experts and contributed a range for concepts of AV-walker interactions. Among them was CyanBand: a horizontal LED light strip on the vehicle's bumper that uses animated cyan lights to communicate AV intentions. Lights sweep from the edges towards the centre when the AV yields and outwards when it does not. CyanBand is the most widely studied eHMI design for walkers \cite{Dey2020TamingInterfaces}, \revised{even appearing in real traffic; the 2024 Mercedes S-Class uses cyan lights on the bumper to communicate that it is in automated mode \cite{Valdes-Dapena2023MercedesSelf-driving}. CyanBand's design was validated in a range of online surveys: \citeauthor{Dey2020ColorPedestrians}~\cite{Dey2020ColorPedestrians} found that while cyan is not as familiar as red or green, it may be more appropriate for communicating AV intentions because it is not used in traffic lights \cite{UK2024TheRules}. This mitigates misinterpretation as crossing instructions (what the pedestrian should do) rather than the AV's intentions (what the AV will do), and avoides liability issues \cite{Anderson2016AutonomousPolicymakers}. Similarly, \citeauthor{Bazilinskyy2019SurveyPerspective}~\cite{Bazilinskyy2019SurveyPerspective} found that placing CyanBand on the bumper yields greater perceived safety than the roof or windscreen, as it is a larger surface. \citeauthor{Colley2024LongitudinalBrowser}~\cite{Colley2024LongitudinalBrowser} compared responses from the USA and Germany to gain a cross-cultural perspective comparing CyanBand to no eHMI. Participants preferred CyanBand across both cultures. These studies provided an early indication of CyanBand's usability for navigating crossings. However, they lack the immersion of life-sized vehicles or moving participants, so these findings may not generalise to AVs in more realistic road scenarios \cite{Al-Taie2025EvARythingInterfaces}.}

Therefore, \citeauthor{Dey2021CommunicatingBehavior}~\cite{Dey2021CommunicatingBehavior} conducted an outdoor study comparing CyanBand with no eHMI. Participants stood in a closed-off crossing, and a vehicle with a hidden human driver approached. They did not physically cross the road for safety reasons. CyanBand outperformed the no eHMI condition, with participants reporting a higher confidence in AV intentions. \revised{However, this study lacked insight into how CyanBand influenced walkers' movement behaviours with the AV, such as whether it required them to slow down or shift their visual attention from the path \cite{Al-Taie2024LightInterfaces}. Participants were also able to interpret the eHMI without the temporal and physical constraints of movement, giving them ample time and cognitive resources to process its signals \cite{Bergstrom-Lehtovirta2011TheInterface}. HCI research already showed how movement can hinder the usability of visual displays \cite{Sarsenbayeva2017ChallengesDevices}. However, most work has focused on smartphones or AR headsets during walking rather than on eHMIs \cite{RubioBaranano2022UsingSpeed, Hoogendoorn2024AObjects, Liu2024TurnAware:Walking}. Differentiating CyanBand’s animations potentially requires pedestrians to continuously monitor the display, yet the multiple resource theory \cite{Wickens2008MultipleWorkload} suggests that mobile users divide attention between interfaces and the surrounding environment to maintain situational awareness \cite{Oulasvirta2005InteractionBursts}. Based on its status as a seminal design in the area and the lack of data on its performance in the demanding scenario of running \cite{Tholander2025TheRunning}, we tested CyanBand in this paper.}

\citeauthor{Al-Taie2024LightInterfaces}~\cite{Al-Taie2024LightInterfaces} conducted a virtual reality (VR) study to evaluate a CyanBand variation (light strips around the vehicle rather than just the bumper) with cyclists. Participants wore a VR headset and pedalled a stationary bike to navigate a virtual environment. They encountered an AV using CyanBand across a range of traffic scenarios, including lane merging and roundabouts. However, participants could not differentiate CyanBand's animations at higher speeds. Consequently, \citeauthor{Al-Taie2024LightInterfaces}~\cite{Al-Taie2024LightInterfaces} developed a new eHMI: LightRing. \revised{This maintained CyanBand's lightstrip form factor, but replaced the cyan animations with colour changes: red flashing quickly meant the AV would not yield, while green pulsing slowly indicated yielding. Follow-up studies consistently showed LightRing to be effective for cyclists because they could distinguish its signals through quick glances while moving. These were conducted outdoors with a hidden human driver \cite{Al-Taie2024LightInterfaces}, across cultures (Sweden, UK, and Oman) \cite{Al-Taie2025AroundCultures}, and when multiple AVs use LightRing around a cyclist \cite{Al-Taie2025EvARythingInterfaces}.} \revised{However, LightRing's use of red and green heightens the risk of these signals being misinterpreted as instructions \cite{Dey2020ColorPedestrians}. Red and green signals were shown to be effective for faster road users \cite{Dey2020TamingInterfaces}, including e-scooter riders \cite{Dey2020TamingInterfaces}, and drivers \cite{Rettenmaier2020AfterScenarios}. However, these road users will encounter AVs in different scenarios from pedestrians, not just at crossings \cite{Berge2024TriangulatingData}. They also move significantly faster using mobility equipment, such as bicycles or scooters \cite{Hollander2021AReview}. Therefore, findings with LightRing may not generalise to pedestrians, the largest road user group \cite{DepartmentforTransport2024ReportedFactsheet, Hollander2021AReview, Dey2020TamingInterfaces}. LightRing still shows particular promise for runners, who move faster than walkers. Consequently, we compared it with CyanBand in this paper to understand how the two most established eHMIs perform with both walkers and runners.} 

\subsection{Runners Provide a New eHMI Use Case}

\revised{Prior work recognised that eHMIs function as public displays that must be inclusive of diverse road user behaviours \cite{Dey2020TamingInterfaces}, and has investigated various road user types \cite{Al-Taie2023KeepTraffic, Matviienko2022E-ScootAR:Reality, Rettenmaier2020AfterScenarios}. \citeauthor{Hollander2021AReview}~\cite{Hollander2021AReview} proposed a taxonomy of the investigated road users to help researchers assess the generalisability of their findings. However, their taxonomy (and nearly all AV interaction research \cite{Dey2020TamingInterfaces}) groups walkers and runners together as pedestrians. While speed is a differentiating factor, research shows that embodied behaviours and movement constraints also fundamentally shape how people use and perceive interfaces \cite{Lepora2015EmbodiedMaking, Kirsh2013EmbodiedDesign, Alexander1996WalkingRunning}. This dimension has been overlooked in eHMI design. Running kinematics differ substantially from walking; in addition to faster movement \cite{Yao2022RecognizingSpeeds} and reduced AV interaction time \cite{Dey2020TamingInterfaces}, runners expend greater physical effort \cite{Mo2017AdaptingMovements}, take longer strides \cite{Alexander1996WalkingRunning}, exhibit less gait stability \cite{Padulo2023GaitSpeeds}, and show higher vertical body motion \cite{Wozniak2015UntanglingRunning}. These embodied differences have already been shown to affect how runners interact with devices, but only in the context of fitness tracking. For example, \citeauthor{Jin2025ResearchScenarios}~\cite{Jin2025ResearchScenarios} asked participants to use a smartwatch to interpret different visualisations of health data (text, charts, graphs) while standing, walking, and running. They found that runners’ increased vertical motion hindered text readability and made animated charts difficult to follow. Runners preferred static icons, raising important questions about the generalisability of animation-based eHMIs, such as CyanBand \cite{Dey2020ColorPedestrians}.}

\revised{Running's physical constraints also narrow the interface design space compared to walking. For example,  \citeauthor{vanRheden2024GestureShirt:Running}~\cite{vanRheden2024GestureShirt:Running} argued that new displays should be integrated into running attire to avoid disrupting movement. They developed GestureShirt by embedding sensors in a running shirt to enable gesture control without breaking running form. Similarly, \citeauthor{Colley2018ShoeDisplays}~\cite{Colley2018ShoeDisplays} introduced shoe-mounted LED displays for runners, proposing their use for tasks such as navigation. These works demonstrate that eHMIs are potentially useful for runners because their on-vehicle placement does not disrupt the current running setup \cite{Berge2022DoStudy}, further motivating our exploration. Similarly, \citeauthor{Seuter2017RunningTechnology}~\cite{Seuter2017RunningTechnology} compared how runners use smartphones, smartwatches, and smart glasses while running on a treadmill. They found that devices must minimise interference with running technique; the smartwatch produced the least disruption because it reduced arm movement and visual attention. In contrast, studies showed that walkers can use less constrained devices, such as smartphones, for complex tasks, such as texting, with little impact on walking movement behaviours \cite{Bergstrom-Lehtovirta2011TheInterface}.} 

\revised{This could be due to situational impairment theory \cite{Marentakis2024SituationalInterfaces}, which suggests that higher physical demand during running temporarily impairs performance in other tasks \cite{Sarsenbayeva2017ChallengesDevices}, potentially influencing how eHMIs should accommodate runners.} \revised{This was also shown in practice: \citeauthor{Epling2018HowMemory}~\cite{Epling2018HowMemory} conducted an outdoor experiment in which participants completed an audio-based narrative memory task while sitting or running. Memory sensitivity declined significantly during running, as participants prioritised maintaining their pace over the cognitive task. In contrast, other studies found that walkers adapt their gait to accommodate a secondary task \cite{Bergstrom-Lehtovirta2011TheInterface, Oulasvirta2005InteractionBursts}; whether conversing with an experimenter on the phone \cite{Murray-Smith2007GaitConversations} or reading AR content \cite{Chang2024ExperienceScenarios}. Therefore, runners prioritised their physical task even if it diminishes their cognitive task performance, while walkers adjusted their movements to accommodate the cognitive task, and this could potentially shape how the two groups respond to eHMIs.} \revised{Embodied cognition theory \cite{Kirsh2013EmbodiedDesign} argues that \textit{``we think with our bodies, not just our brains''}. This suggests that pedestrians integrate their physical constraints and bodily cues into decision-making, not just visual display feedback. For example, \citeauthor{Tholander2025TheRunning}~\cite{Tholander2025TheRunning} conducted an autoethnographic study during running. They showed that, despite fitness tracker feedback on heart rate and running pace, runners also rely on bodily cues for fatigue and emotional state to assess and verify the tracker's information. Therefore, the distinct physical experiences of walking and running could determine how pedestrians use their bodily feedback to interpret eHMI information.} 

\revised{Despite this, no prior work investigated AV-runner interactions \cite{Hollander2021AReview, Dey2020TamingInterfaces, Rasouli2020AutonomousPractice, HegnaBerge2023SupportOutlook}. Very little research explored how runners perceive and behave around vehicles. Survey studies by \citeauthor{Ettema2016RunnableFrequency}~\cite{Ettema2016RunnableFrequency} and \citeauthor{Deelen2019AttractiveEnvironment.}~\cite{Deelen2019AttractiveEnvironment.} show that runners identify vehicles as major obstacles because they prefer to maintain their pace, aligning with \citeauthor{Epling2018HowMemory}'s~\cite{Epling2018HowMemory} findings. This contrasts with real-world observations showing that walkers often stop at crossings \cite{Haddington2014InteractionTraffic, Dey2017TheVehicles}. Together, these findings highlight that runners exhibit distinct behaviours around vehicles and must be studied independently to ensure eHMIs align with their needs \cite{Hollander2021AReview}, a task we pick up in this paper.} 

\subsection{Study Designs for eHMI Evaluation}

\revised{Pedestrian embodied behaviours can shape how they use eHMIs \cite{Kirsh2013EmbodiedDesign}, so it is crucial to test displays using study designs that trigger realistic behaviours to reflect real-world interactions.} One approach is \citeauthor{Rothenbucher2015GhostDriver}'s~\cite{Rothenbucher2015GhostDriver} Ghost Driver \cite{Mahadevan2018CommunicatingInteraction, Al-Taie2024LightInterfaces, Dey2021CommunicatingBehavior}. This involves a hidden human driver in a real vehicle to simulate an AV. However, this makes it unsafe to include non-yielding AV behaviours. Previous studies overcame this by only testing yielding AVs \cite{Al-Taie2024LightInterfaces} or having participants remain stationary \cite{Dey2021CommunicatingBehavior}. However, these approaches either do not expose participants to all eHMI states or do not collect participant movement behaviours. Therefore, follow-up AV-walker interaction studies used VR simulators \cite{Dey2021TowardsPedestrian, Dey2021TowardsPedestrian}. Participants could navigate a fully virtual crossing by walking indoors to experience all AV-yielding behaviours without real danger \cite{Tran2021AInteraction}. \revised{However, VR studies require a treadmill or indoor environment to support participant safety, as they cannot see the real world around them \cite{Tran2021AInteraction}. This reduces the sense of real urban walking and running \cite{Dey2021TowardsPedestrian, Al-Taie2024BikeInterfaces}.}

As AR and passthrough features in mixed-reality headsets have advanced, recent research used AR simulators to project virtual objects onto outdoor space. This allows participants to navigate real outdoor space while interacting with life-sized virtual objects \cite{Matviienko2022BikeAR:Reality, Matviienko2022E-ScootAR:Reality}. \citeauthor{Aleva2024AugmentedStudy}~\cite{Aleva2024AugmentedStudy} used this approach to test various AR-based interfaces for AV-walker interaction, and \citeauthor{Al-Taie2025EvARythingInterfaces}~\cite{Al-Taie2025EvARythingInterfaces} developed an open-source AR simulator, CycleARcade\footnote{CycleARcade: An open-source AR simulator to test AV-road user encounters - \url{github.com/ammarjamal/CycleARcade} (Accessed: 31/08/2025)}, and used it to test AV interfaces across multiple studies \cite{Al-Taie2025AroundCultures, Al-Taie2025EvARythingInterfaces}. \revised{All studies consistently showed that this AR approach triggers realistic road user behaviours, with participants reporting a high sense of presence and immersion, and a realistic sense of risk and safety \cite{Matviienko2022BikeAR:Reality, Matviienko2022E-ScootAR:Reality, Al-Taie2025EvARythingInterfaces, Aleva2024AugmentedStudy}.} Based on these effective demonstrations, we adapted CycleARcade for pedestrian use in this paper. More recently, \citeauthor{Abe2025IRunners}~\cite{Abe2025IRunners} developed RunSight an AR tool to help low-vision individuals run at night. RunSight was deployed on a Meta Quest 3 headset\footnote{Meta Quest 3 mixed-reality headset: \url{meta.com/quest/quest-3} (Accessed: 31/08/2025)} and tested in a real-world study. The authors found that low-vision participants were able to run at least 1km with RunSight. This further encouraged us to utilise an AR simulator in our study, as it showed that runners can use such systems while moving in physical space.

\subsection{Summary and Research Questions}

eHMIs are promising solutions for AV-pedestrian interaction \cite{Dey2019GazeStudies, Dey2020TamingInterfaces}. However, they are public displays that must be inclusive of road user behaviours \cite{Hollander2021AReview}. While the research expanded to other road user types, such as cyclists and drivers \cite{HegnaBerge2023SupportOutlook, Hollander2021AReview}, pedestrian research remains focused on walking speeds \cite{Rasouli2020AutonomousPractice}. It is unknown how eHMIs can be designed to support runners, who are a significant subset of the pedestrian user group \cite{RunningWithGrit2025StrikingLife}. Runners move faster and experience distinct bodily demands from walkers \cite{Alexander1996WalkingRunning}. \revised{Theories on situational impairment \cite{Marentakis2024SituationalInterfaces}, attentional demand \cite{Wickens2008MultipleWorkload}, and embodied cognition \cite{Kirsh2013EmbodiedDesign} suggest that these differences shape how runners use and perceive eHMIs, questioning the generalisability of walker-centric designs.} Therefore, we ask the following research questions (RQs) to understand how the distinct movements of walkers and runners influence their AV interactions, and assess the generalisability of established eHMIs (LightRing \cite{Al-Taie2024LightInterfaces} and CyanBand \cite{Dey2020ColorPedestrians}) to promote inclusivity between these road users:

\begin{itemize}
    \item[\textbf{RQ1}] \revised{How do the bodily and perceptual demands of walking and running shape pedestrians’ movement behaviours (e.g., speed adjustments and head rotations) when responding to eHMIs during AV encounters?}
    \item[\textbf{RQ2}] \revised{What effects do these movement behaviours have on an eHMI's perceived ability to support trust, safety, workload and usability in AV interactions?}
\end{itemize}
\section{Methodology}

To address the RQs, we conducted a within-subjects study in which participants used an AR simulator to navigate a virtual crossing while either walking or running. An approaching AV either yielded or did not yield. It displayed one of three eHMI conditions to communicate its intentions: LightRing \cite{Al-Taie2024LightInterfaces} (colour-changing red/green lights around the vehicle), CyanBand \cite{Dey2020ColorPedestrians} (animated cyan lights on the vehicle’s front), or a No-eHMI baseline. We measured participant crossing behaviours toward each eHMI via simulator logs and perceptions through questionnaires. For the first time, we show how pedestrian speed and embodied behaviours influence eHMI use to support inclusive AV interfaces.

\subsection{Apparatus and AR Simulator Development}
\begin{figure*}
    \centering
    \includegraphics[width=\linewidth]{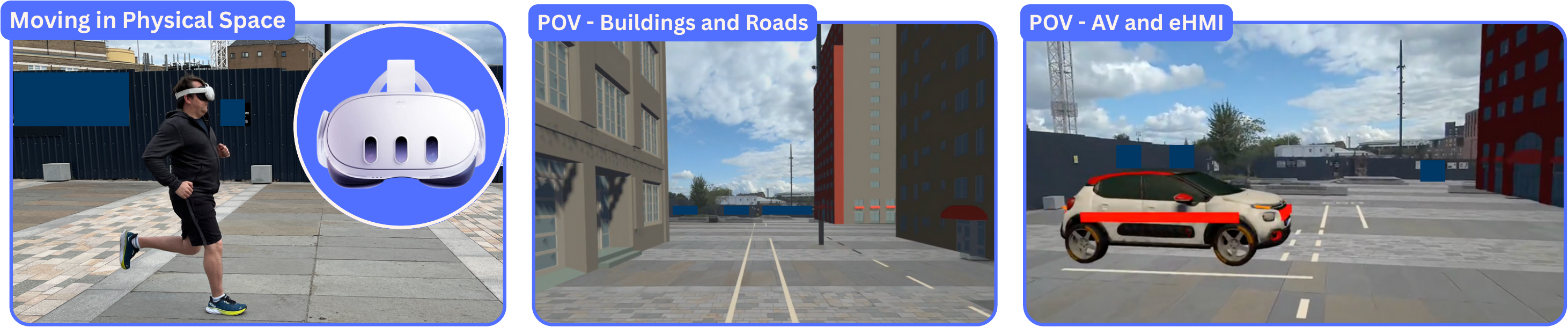}
    \caption{The AR simulator used in the study. The left image shows a participant moving in real physical space while wearing a Meta Quest 3 headset. The middle image shows the participant's point of view (POV) of the virtual urban environment projected by the headset with passthrough, and the right shows the virtual AV with the LightRing eHMI approaching the crossing and not yielding.}
    \label{fig:apparatus}
\end{figure*}

\revised{Our study design required participants to experience eHMIs across yielding and non-yielding AV behaviours. This prohibited us from using a real vehicle, as it would be unsafe for participants to cross when the vehicle does not yield \cite{Rothenbucher2015GhostDriver, Dey2021TowardsPedestrian, Al-Taie2024LightInterfaces}. Therefore,} we adapted CycleARcade \cite{Al-Taie2025EvARythingInterfaces}, an open-source AR simulator originally developed for cyclists, to pedestrian use. We named our variation ARcade for simplicity; see \autoref{fig:apparatus}. ARcade projected a virtual urban environment onto a real outdoor space, including buildings, roads, AVs and \revised{high-fidelity} eHMIs. This allowed participants to move freely in the real world while interacting with to-scale virtual objects, crucial for capturing natural walking and running behaviours \cite{Matviienko2022BikeAR:Reality, Aleva2024AugmentedStudy}. This approach was used in prior work \cite{Al-Taie2025EvARythingInterfaces, Al-Taie2025AroundCultures} and shown to immerse participants, triggering realistic road user behaviours. CycleARcade was developed in Unity \cite{Al-Taie2025EvARythingInterfaces};  adapting it involved replacing any virtual cycle lanes with pedestrian pavements and implementing a road crossing scenario.

All AVs were SAE Level 5 automated vehicles \cite{SocietyforAutomotiveEngineers2021SAEVehicles} with no passengers. They were 3D models of a Citroën C3, a common city car. AVs followed pre-recorded driving animations, useful for maintaining consistent driving behaviours throughout the study \cite{Al-Taie2025AroundCultures}. We deployed ARcade on a Meta Quest 3 mixed-reality headset. This was used to test AR displays for runners in prior work \cite{Abe2025IRunners}. It supports coloured passthrough and depth-sensing to provide a clear real-world background and accurate alignment of opaque virtual objects with the physical environment, enhancing immersion. \cite{Al-Taie2025AroundCultures}. All questionnaires were online forms that participants answered using a tablet.

\subsection{Assessed eHMIs}
\label{ehmi}

\begin{figure*}
    \centering
    \includegraphics[width=\linewidth]{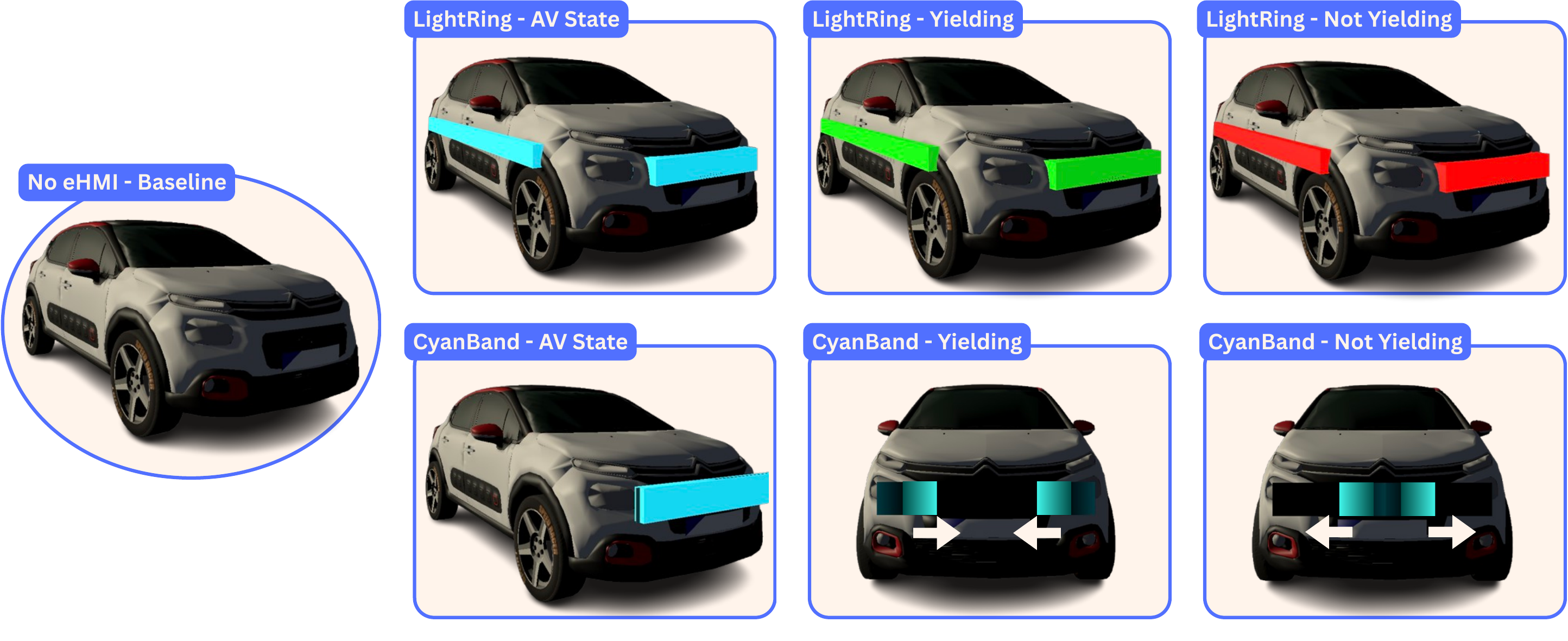}
    \caption{The eHMIs explored in the study. The left image shows the baseline condition with No-eHMI. Top: LightRing communicating its autonomous state (vehicle is automated with all sensors functioning), its yielding state and then its non-yielding state. Bottom: CyanBand communicating the same three states; the arrows show the direction of cyan animations. }
    \label{fig:ehmi}
\end{figure*}

We tested three eHMI conditions: LightRing \cite{Al-Taie2024LightInterfaces}, CyanBand \cite{Dey2018InterfaceUsers}, and No-eHMI; see \autoref{fig:ehmi}. \revised{The eHMIs were adapted from prior work \cite{Dey2020TamingInterfaces, Al-Taie2024LightInterfaces} to evaluate whether established designs generalise to runners or if new ones are needed. This allowed us to prioritise eHMI inclusivity \cite{Hollander2021AReview}, rather than develop runner-specific concepts. Each eHMI is the most mature design for the road user type it was developed for, having been refined and iterated across multiple studies \cite{Dey2020TamingInterfaces, Al-Taie2025EvARythingInterfaces, Al-Taie2025AroundCultures, Dey2020ColorPedestrians, Dey2021CommunicatingBehavior}, and even appearing in real traffic \cite{Valdes-Dapena2023MercedesSelf-driving}. LightRing was developed for cyclists \cite{Al-Taie2024BikeInterfaces, Al-Taie2024LightInterfaces, Al-Taie2025AroundCultures, Al-Taie2025EvARythingInterfaces}, while CyanBand was developed for walkers \cite{Dey2018InterfaceUsers, Dey2020ColorPedestrians, Dey2021CommunicatingBehavior, Dey2024Multi-ModalInteraction}. This allowed us to position runners within the existing design space of road users moving at different speeds \cite{Hollander2021AReview, Rasouli2020AutonomousPractice, Sandt2017DiscussionBicyclists}, and verify whether additional factors, such as physical constraints and movement behaviours \cite{Kirsh2013EmbodiedDesign}, should be used to differentiate road users.}  The eHMIs used only visual cues because auditory signals may be lost in environmental noise and are challenging to associate with the AV \cite{Al-Taie2023PimpCyclists, Dey2024Multi-ModalInteraction}. Consistent with previous studies \cite{Al-Taie2024LightInterfaces, Al-Taie2025EvARythingInterfaces, Al-Taie2025AroundCultures}, all eHMIs began responding to the pedestrian when the AV was 15m away. They worked as follows:

\begin{itemize}
    \item \textbf{LightRing:} This eHMI consists of LED light strips positioned around the vehicle’s body, including the front, sides, and rear, to support broader visibility. The lights remain on in cyan to indicate that the vehicle is in automated mode with all sensors functioning properly. When the vehicle intends to yield, the lights pulse slowly in green (1 pulse per second); when it does not intend to yield, they flash rapidly in red (2 flashes per second). These pulsing/flashing animations were added to support colour-blind road users \cite{Al-Taie2024LightInterfaces}. \revised{Road users could potentially distinguish LightRing's colour-based signals through quick glances, supporting faster crossing decisions \cite{Al-Taie2024LightInterfaces}. However, it was only tested with cyclists \cite{Al-Taie2024BikeInterfaces, Al-Taie2025AroundCultures}, so it may entail limitations with walkers and runners, such as red and green being misinterpreted as instructions \cite{Dey2020ColorPedestrians}, prompting us to verify its inclusivity.}
    
    \item \textbf{CyanBand:} A horizontal LED light strip mounted on the vehicle's front bumper, directly facing pedestrians at crossings \cite{Dey2019GazeVehicles}. Like LightRing, the lights remain on in cyan to indicate that the vehicle is in automated mode. However, instead of colour, CyanBand conveys AV intentions through directional animation. When the vehicle intends to yield, the cyan lights continuously sweep inward from the edges toward the centre at a rate of 1 sweep per second. When the vehicle does not intend to yield, the lights continuously sweep outward from the centre to the edges at the same rate. \revised{It was crucial to test CyanBand with moving participants because directional animations may require continuous monitoring \cite{Al-Taie2024LightInterfaces}, which contradicts HCI theories that mobile users allocate attention to displays for only seconds at a time \cite{Wickens2008MultipleWorkload, Oulasvirta2005InteractionBursts}. Unlike LightRing, CyanBand is placed only on the vehicle's front, allowing us to investigate both the eHMI placement and visual cue type (animation vs. colour changes) as design considerations \cite{Al-Taie2023KeepTraffic}.}

    \item \textbf{No-eHMI:} This was a baseline condition with no eHMI display on the AV. Instead, vehicle intentions had to be inferred through implicit signals in the AV's driving and braking behaviours \cite{Markkula2020DefiningTraffic}. Previous research showed that walkers can still make safe crossing decisions without the support of eHMIs \cite{Moore2019TheVehicles, Dey2017TheVehicles, Lee2021RoadVehicles}. However, it is unclear whether this generalises to runners.
    \end{itemize}

\subsection{Developing a Crossing Scenario}
\label{scenario}

\begin{figure*}
    \centering
    \includegraphics[width=0.75\linewidth]{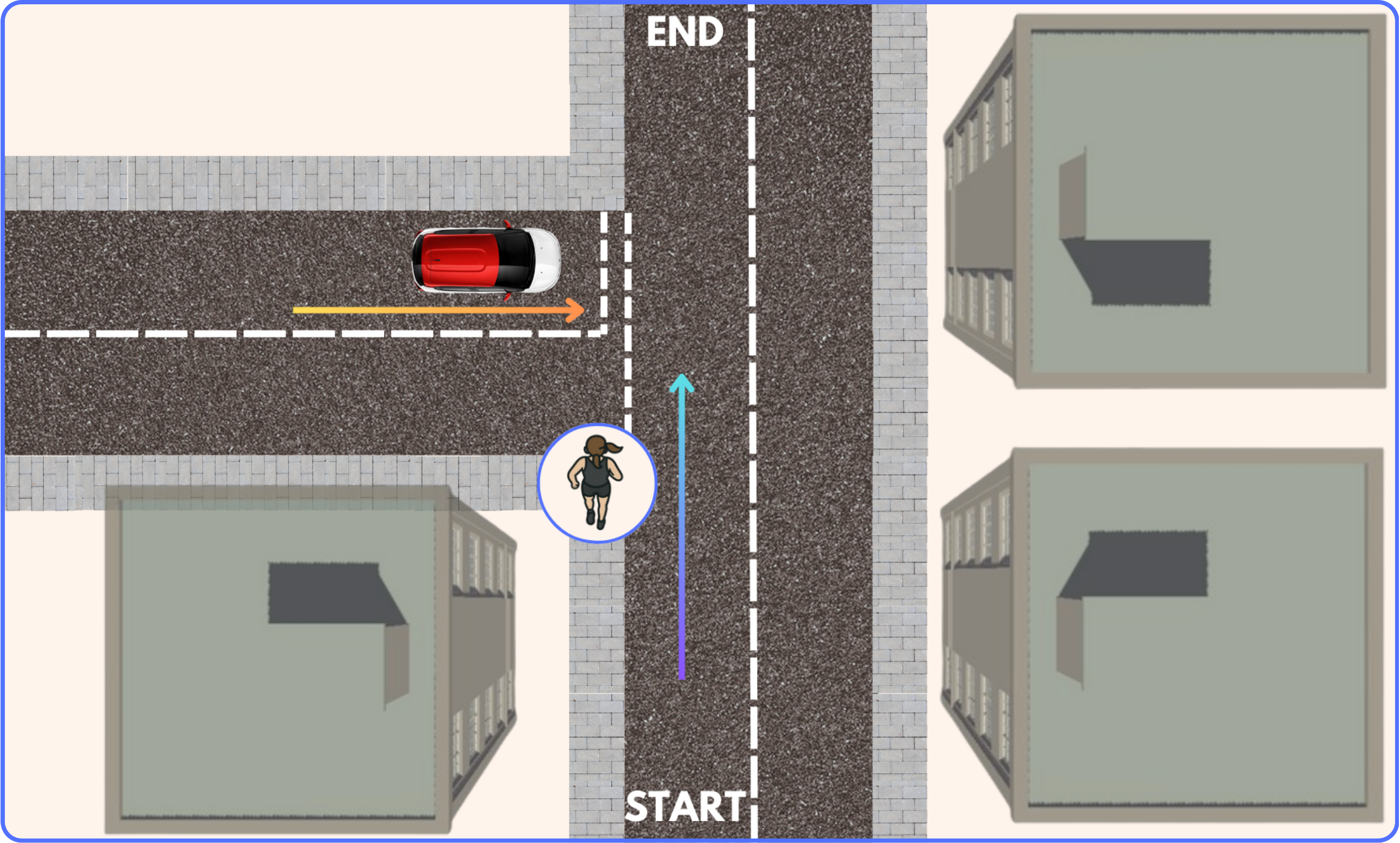}
    \caption{Top-down view of the study's crossing scenario. A two-lane intersection with one lane for an AV turning left onto the main road and another for a vehicle turning off the main road. The AV approached from the pedestrian’s left, travelling in the lane furthest from them. The blue arrow indicates the pedestrian’s path; the orange arrow indicates the AV’s path.}
    \label{fig:scenario}
\end{figure*}

We followed a systematic approach to design a crossing scenario in ARcade that reflects typical vehicle-runner encounters in the UK (see \autoref{fig:scenario}), \revised{thereby strengthening the broader applicability and ecological validity of our setup \cite{Al-Taie2024LightInterfaces}}. First, we checked the Strava heatmap\footnote{Strava heatmap: Shows running traffic from Strava users, a common fitness tracking application; \url{strava.com/maps/global-heatmap} (Accessed: 31/08/2025)} to identify the five most popular running routes near the University. All routes included at least one two-lane intersection: one lane for vehicles entering from a side road onto the main road and another for vehicles exiting the main road. The intersections had no traffic lights to regulate right-of-way, requiring communication between road users \cite{Dey2017TheVehicles, Markkula2020DefiningTraffic}. \revised{Following this, we checked the UK Highway Code \cite{UK2024TheRules} and national surveys of urban junctions \cite{ActiveTravel2024Cycling2022, DepartmentforTransport2024ReportedFactsheet} to ensure that this intersection type extends beyond our local area. This is the most common intersection layout across the UK, supporting the external validity of our setup \cite{Berge2024TriangulatingData}. Unsignalised intersections are also the most common scenario in AV interaction research \cite{Dey2020TamingInterfaces}; they have been explored with walkers, cyclists, and drivers \cite{Rasouli2020AutonomousPractice, Berge2024TriangulatingData, Dey2020TamingInterfaces}, facilitating clearer comparability of our results \cite{Hollander2021AReview}.} Next, we visited five of the intersections identified on Strava to document key features, such as road markings. Our observations were consistent with the UK Highway Code \cite{UK2024TheRules}: lane widths ranged between 3.5m and 4.0m, so we adopted the 3.65m UK standard. Side roads leading to a main road featured give-way lines marked as two dashed white lines, while lanes exiting the main road had give-way line markings with a single dashed line. These features were incorporated into the virtual scenario \revised{to ensure the setup reflected real-world layouts and supported the generalisability of our outcomes \cite{Al-Taie2023KeepTraffic, Berge2024TriangulatingData}.}

The AV drove at 48 km/h, the UK standard for urban areas, \revised{to facilitate representative driving behaviours \cite{Al-Taie2023KeepTraffic, Al-Taie2024LightInterfaces}}. It approached from the side road on the pedestrian’s left, driving in the lane furthest from the pedestrian. This added ambiguity and urgency to the crossing decision because the pedestrian may have already entered the road (in the first lane) as the AV approached. It also made it more realistic to include non-yielding AV scenarios, because the pedestrian would not be directly opposite the AV when starting to cross \cite{Al-Taie2025AroundCultures}. When yielding, the AV stopped 0.5m behind the give-way line; when not yielding, it maintained speed and turned onto the main road.

\subsection{Study Design}
The experiment used a within-subjects design with three independent variables: pedestrian \textit{activity} (walking vs. running), \textit{AV-behaviour} (yielding, not-yielding) and \textit{eHMI} (LightRing, CyanBand and No-eHMI). The dependent variables were participant  \textit{behaviours} and \textit{perceptions} toward each eHMI. The study took place in a quiet, pedestrianised outdoor space measuring 50m × 40m, free of real traffic to ensure participant safety \cite{Aleva2024AugmentedStudy}. We investigated how pedestrian activity influenced eHMI use. However, it was crucial also to include AV-behaviour as an independent variable to expose participants to all eHMI states and examine usability when the AV slowed to yield and its behaviour clearly aligned with the eHMI signal, versus when it maintained speed and participants had to interpret the eHMI signal more quickly. This also reduced learning effects, as participants were unsure whether the AV would stop at the crossing \cite{Al-Taie2024LightInterfaces}. Participants used ARcade to navigate the crossing scenario described in \autoref{scenario} and experienced all study conditions. Each participant completed 12 trials to test every eHMI while walking and running under both AV-yielding behaviours. Conditions were counterbalanced using a Latin square design to ensure all participants experienced each combination and to minimise ordering effects. 

The ARcade environment was marked with start and endpoints. In each trial, participants walked or ran 30m from the start to the crossing, allowing them to reach their standard pace. They continued 7.3m through the crossing (where the AV approached), and travelled an additional 5m to the end. This final segment encouraged participants to maintain their pace beyond the crossing, important for capturing realistic running behaviour \cite{Deelen2019AttractiveEnvironment.}. In ARcade, the AV was triggered to move once the participant reached a predetermined location approaching the crossing. For runners, this was set 2m earlier than for walkers to avoid the late arrival of the AV. Trigger locations were determined through ten pilot tests to ensure the AV and participant reached the crossing simultaneously, important for forming a space-sharing conflict \cite{Markkula2020DefiningTraffic}.

\subsection{Measures}

\revised{To answer the RQs, we measured the following pedestrian movement behaviours and perceptions toward each eHMI:}

\begin{enumerate}
    \item \textbf{\revised{Movement and Crossing Behaviours:}} \revised{To answer RQ1, ARcade logged movement behaviours that reflected the physical ways walkers and runners interacted with the AV across eHMI conditions. Participant speed (m/s) was recorded every 100 milliseconds to understand whether participants would maintain their pace or slow down to interpret eHMI signals \cite{Deelen2019AttractiveEnvironment.}. ARcade also recorded the number of times participants stopped at the crossing, defined as a speed below 0.2 m/s, based on 10 pilot tests; stopping requires more significant movement changes than slowing down \cite{Yao2022RecognizingSpeeds}, but could give more time to interpret eHMI signals and diminish any body movements that potentially hinder comprehensibility \cite{Alexander1996WalkingRunning}. Therefore, this measure showed whether there was a tradeoff between movement rhythm and decision-making time. We also collected participant head angles (Unity camera Y-axis rotation, logged every 100 milliseconds) to indicate the extent to which participants diverted attention from the path ahead to check AV intentions \cite{Al-Taie2025AroundCultures}; this was important because the eHMIs were placed differently, around the vehicle versus on the front. It also showed whether directing visual attention toward the AV or eHMI impacted other movement behaviours. Finally, to explore the effects of these movements, ARcade recorded any collisions between participants and the AV;}
    
    \item \textbf{\revised{eHMI Impact on the Crossing Experience:}} \revised{To address RQ2, we administered a post-crossing questionnaire after each trial to examine how movement behaviours from the preceding measures influenced participants' eHMI use. This comprised six five-point Likert-scale questions (strongly disagree to strongly agree) informed by prior AV interaction research \cite{Al-Taie2024LightInterfaces, Dey2021CommunicatingBehavior, Dey2020ColorPedestrians, Al-Taie2025AroundCultures}, alongside standardised measures assessing workload and usability. First, participants indicated the eHMI's role in their crossing decision: \textit{``I relied more on the eHMI than the AV’s driving behaviour''} and \textit{``The AV instructed me to cross rather than communicating its intentions''}. eHMIs should support informed decisions based on a combination of implicit and explicit cues \cite{Markkula2020DefiningTraffic}, and facilitate negotiation by communicating the AV's intended actions, rather than instructing pedestrians \cite{Dey2020ColorPedestrians}.} \revised{Second, participants indicated how the eHMI shaped trust and confidence: \textit{``I trusted the AV to yield/not yield''} and \textit{``I was confident in the AV’s intentions''}. High trust and confidence require pedestrians to align eHMI signals with the vehicle's implicit behaviours \cite{Al-Taie2025EvARythingInterfaces, Chang2017EyesPedestrian}, which may be challenging given the temporal and physical constraints during movement \cite{Alexander1996WalkingRunning}. Third, the questionnaire assessed the implications of participant crossing decisions on risk and safety: \textit{``My crossing decision was risky''} and \textit{ ``I felt safe crossing the road''}. Effective eHMIs must minimise risk and heighten perceived safety \cite{Rasouli2020AutonomousPractice}; however, it remains unknown how these effects generalise between walking and running. After answering the Likert-scale questions, participants completed the NASA Task Load Index (NASA-TLX) \cite{Hart1988DevelopmentResearch} to indicate the workload to safely cross and reach the endpoint without collision, and the User Experience Questionnaire-Short Version (UEQ-S) \cite{Schrepp2017DesignUEQ-S} to report the usability of the AV interaction. Finally, participants provided free-text feedback for further reflections and context. Inclusive eHMIs should achieve consistent scores across this set of measures \cite{Dey2020TamingInterfaces, Al-Taie2024LightInterfaces}. The full questionnaire is in the supplementary materials;}
    
    \item \textbf{\revised{Qualitative Insights on the Crossing Experience:}} After completing all study conditions, participants ranked the eHMIs from best to worst overall, and then separately for walking and running. This showed whether walkers and runners preferred different eHMIs. Participants also provided free-text responses, offering qualitative insights into the eHMIs and perceived differences between crossing while walking versus running.
\end{enumerate}

\subsection{Participants}

The study involved 24 participants (13 males, 11 females;  mean age = 25 years, SD = 5.6; \revised{age range = 18-61 years}). They were recruited through \revised{university mailing lists, reaching a broad pool of students, staff, affiliates, and general public who had subscribed to study notifications. In accordance with institutional ethical guidelines, individuals with no running experience were not recruited to avoid risk of overexertion or injury. Therefore, we included participants who engaged in running either regularly or occasionally within the past five years, ensuring the sample represented varying levels of running experience similar to those likely to encounter AVs: Fifteen participants reported running multiple times per week, five multiple times per month, and four multiple times per year.}

\subsection{Procedure}

Each participant met the experimenter at the study location. They were briefed about the study and completed a demographic survey. The experimenter then calibrated the headset and aligned the AR view with the real-world floor level. The participant put on the headset and practised walking in the virtual environment from the start to the endpoint without an approaching AV. This mitigated novelty effects and familiarised the participant with the setup. The participant then removed the headset, and the experimenter introduced the eHMIs using photographs of the displays.

The main experiment then commenced. Based on a Latin square design, the experimenter selected a running/walking condition, AV-yielding type, and eHMI via a menu in ARcade. The participant put on the headset again and was instructed to either walk or run from the start to the endpoint; they were asked to behave as they normally would. \revised{The experimenter observed the participant to ensure they walked or ran as instructed.} An AV approached the crossing while displaying the selected eHMI; it either yielded or did not yield. Upon reaching the endpoint, the participant removed the headset and completed the post-crossing questionnaire on a tablet. This process was repeated 12 times to ensure exposure to all conditions. At the end of the session, the participant used an online form to rank the eHMIs and explain their preferences. The study lasted approximately 45 minutes and was approved by the University’s ethics committee. Each participant was compensated for their time with a £10 Amazon voucher.

\section{Results}

This section shows how walkers and runners adjusted their movement behaviours between eHMI conditions, and how these behaviours influenced their perceptions and use of the eHMIs. It also reports qualitative themes from participant feedback. The raw data are available at \url{zenodo.org/records/18344381}.

\begin{table*}[]
\footnotesize
\begin{tabular}{@{}llllllll@{}}
\toprule
                                                       & \multicolumn{2}{c}{\textbf{Activity}}                    & \multicolumn{3}{c}{\textbf{eHMI}}                                              & \multicolumn{2}{c}{\textbf{AV-behaviour}} \\ \midrule
\multicolumn{1}{l|}{}                                  & \textit{Walking} & \multicolumn{1}{l|}{\textit{Running}} & \textit{LightRing} & \textit{CyanBand} & \multicolumn{1}{l|}{\textit{No-eHMI}} & \textit{Yielding}      & \textit{Not-Yielding}     \\ \midrule

\multicolumn{1}{l|}{\textbf{Crossing Speed}}           & 1.75 ± 0.58      & \multicolumn{1}{l|}{2.62 ± 0.63}      & 2.64 ± 0.76        & 1.95 ± 0.60       & \multicolumn{1}{l|}{1.98 ± 0.65}      & 2.17 ± 0.69            & 2.23 ± 0.80               \\
\multicolumn{1}{l|}{\textbf{Head Rotations}}           & 15.97 ± 9.32     & \multicolumn{1}{l|}{19.23 ± 9.03}     & 20.61 ± 11.74      & 16.15 ± 7.44      & \multicolumn{1}{l|}{16.06 ± 7.15}     & 17.84 ± 8.59           & 17.48 ± 9.98              \\ \midrule
\multicolumn{1}{l|}{\textbf{Reliance on eHMI}}         & 3.08 ± 1.63      & \multicolumn{1}{l|}{3.66 ± 1.46}      & 4.26 ± 1.06        & 3.08 ± 1.57       & \multicolumn{1}{l|}{-}                & 3.72 ± 1.39            & 3.62 ± 1.54               \\
\multicolumn{1}{l|}{\textbf{Perceived Instruction}}    & 3.15 ± 1.44      & \multicolumn{1}{l|}{2.92 ± 1.49}      & 4.09 ± 1.14        & 2.81 ± 1.30       & \multicolumn{1}{l|}{2.19 ± 1.26}      & 2.85 ± 1.51           & 3.21 ± 1.40              \\
\multicolumn{1}{l|}{\textbf{Trust}}                    & 3.43 ± 1.36      & \multicolumn{1}{l|}{3.03 ± 1.46}      & 4.26 ± 1.05        & 3.17 ± 1.25       & \multicolumn{1}{l|}{2.26 ± 1.19}      & 3.36 ± 1.34            & 3.10 ± 1.50               \\
\multicolumn{1}{l|}{\textbf{AV Intention Confidence}} & 3.44 ± 1.38      & \multicolumn{1}{l|}{3.01 ± 1.46}      & 4.26 ± 1.11        & 3.11 ± 1.22       & \multicolumn{1}{l|}{2.30 ± 1.25}      & 3.30 ± 1.37            & 3.15 ± 1.50               \\
\multicolumn{1}{l|}{\textbf{Perceived Risk}}           & 2.15 ± 1.21      & \multicolumn{1}{l|}{2.67 ± 1.36}      & 1.82 ± 1.01        & 2.31 ± 1.18       & \multicolumn{1}{l|}{3.09 ± 1.40}      & 2.43 ± 1.29            & 2.39 ± 1.33               \\
\multicolumn{1}{l|}{\textbf{Perceived Safety}}         & 3.88 ± 1.21      & \multicolumn{1}{l|}{3.26 ± 1.37}      & 4.30 ± 0.97        & 3.51 ± 1.23       & \multicolumn{1}{l|}{2.90 ± 1.36}      & 3.63 ± 1.23            & 3.51 ± 1.42               \\
\multicolumn{1}{l|}{\textbf{Crossing Workload}}        & 2.70 ± 2.06      & \multicolumn{1}{l|}{5.08 ± 2.66}      & 3.09 ± 2.13        & 4.00 ± 2.71       & \multicolumn{1}{l|}{4.58 ± 2.88}      & 3.82 ± 2.57            & 3.96 ± 2.74               \\
\multicolumn{1}{l|}{\textbf{Usability}}         & 0.35 ± 0.96      & \multicolumn{1}{l|}{0.29 ± 0.96}      & 0.94 ± 0.76        & 0.09 ± 0.93       & \multicolumn{1}{l|}{-0.06 ± 0.86}     & 0.40 ± 0.94            & 0.24 ± 0.98               \\

\bottomrule
\end{tabular}
\caption{Mean ± SD of the behavioural measures and post-crossing questionnaire subscales per independent variable. Reliance on eHMI is omitted for the No-eHMI condition, as there was no display.}
\label{tab:post-crossing-mean}
\end{table*}

\subsection{\revised{Movement and Crossing Behaviours}}

This section reports results on walker and runner movement behaviours in response to each eHMI, as logged by ARcade. Mean values for \textit{Crossing Speed} and \textit{Head Rotations} are in \autoref{tab:post-crossing-mean}. The data for these two metrics were not normally distributed. Therefore, we conducted a three-way Aligned Rank Transform (ART) ANOVA \cite{Wobbrock2011TheProcedures} to examine the fixed effects of \textit{pedestrian activity}, \textit{AV-behaviour}, \textit{eHMI}, and their \textit{interactions} on \textit{Crossing Speed} and \textit{Head Rotations}. The model included a random intercept for \textit{participant} to account for individual variability. \textit{Post hoc} comparisons were performed using the ART-C method \cite{Elkin2021AnTests} for significant factors with more than two levels (\textit{eHMI} and the \textit{interactions}). We report only significant interactions for clarity; the full results are in the supplementary materials. \autoref{fig:behaviour} also visualises the logged movement behaviours for each eHMI per walking and running conditions.

\paragraph{\textbf{Crossing Speed}}  There was a significant main effect of \textit{activity} (F(1, 244) = 362.31, $P< .001$; $\eta^{2}$ = 0.6), showing that \textit{runners} were faster than \textit{walkers}. \revised{This acted as a post-study manipulation check, confirming that participants followed the running/walking task as instructed.} There was no effect of \textit{AV-behaviour} (F(1, 243) = 1.33, $P=.3$), meaning participant speeds remained consistent regardless of whether the AV yielded. There was a significant main effect of \textit{eHMI} (F(2, 243) = 71.57, $P< .001$; $\eta^{2}$ = 0.37), and \textit{post hoc} tests showed that participants were significantly faster with \textit{LightRing} than the other conditions ($p<.0001$ for both).  

\paragraph{\underline{Interactions}} There was a significant interaction between \textit{activity and eHMI} (F(2, 243) = 4.57, $P=.01$; $\eta^{2}$ = 0.04), meaning that each eHMI's influence on pedestrian speed differed between walking and running. \textit{Post hoc} tests showed that participants were significantly faster using \textit{LightRing while running} than all other conditions ($p < .005$ for all). In contrast, they were significantly slower using \textit{CyanBand while walking} than all other conditions ($p < .0001$ for all).

\paragraph{\textbf{Stopping Frequency}} Participants stopped 163 times while \textit{walking} (73 times with a yielding AV and 90 with a non-yielding AV), and 63 times while \textit{running} (23 with a yielding AV and 40 with a non-yielding AV). We conducted chi-square tests of independence to examine the relationships between our independent variables on the likelihood of stopping. \textit{Post hoc} comparisons were conducted using a chi-square test of independence with a Bonferroni correction. 

There was no association between \textit{activity} and \textit{AV-behaviour} ($\chi^2$(1, 226) = 0.143, $p = .7$), so participants were more likely to stop when the AV did not yield, regardless of their activity. However, there was a significant association between \textit{activity} and \textit{eHMI} ($\chi^2$(2, 226) = 19.596, $p < .001$); the effect of an eHMI on stopping depended on whether participants were walking or running. \textit{Post hoc} comparisons showed that \textit{walkers} were less likely to stop with \textit{LightRing} than the other eHMIs ($p<.001$ for both). In contrast, \textit{runners} were less likely to stop with both \textit{LightRing} and \textit{No-eHMI} than \textit{CyanBand} ($p<.001$ for both). Therefore, directional animations required participants to stop more frequently than implicit driving behaviours, suggesting they \revised{hindered pace and were difficult to process whilst in motion.}

\paragraph{\textbf{Head Rotations}} There was a significant main effect of \textit{activity} (F(1, 244) = 19.59, $P< .001$; $\eta^{2}$ = 0.07); \textit{runners} made larger head rotations than \textit{walkers}, \revised{suggesting that runners compensated for not stopping by continuously checking the AV while moving, but walkers waited for the AV to be in their view.} There was no effect of \textit{AV-behaviour} (F(1, 243) = 1.23, $P=.3$). There was a significant main effect of \textit{eHMI} (F(2, 243) = 8.64, $P< .001$; $\eta^{2}$ = 0.07). \textit{Post hoc} tests showed that rotations were significantly larger with \textit{LightRing} than the others ($p<.005$ for both), potentially because participants \revised{were least likely to stop when using LightRing, so they continuously checked the eHMI}. There were no interactions.

\paragraph{\textbf{Collisions}} Three collisions occurred, all while running, when the AV did not yield. One happened when there was \textit{No-eHMI}, and two with \textit{LightRing}. The No-eHMI collision occurred because the participant misinterpreted the AV's yielding intentions:  \textit{``There was no display. I misinterpreted what it would do''} - P4. In contrast, collisions involving \textit{LightRing} happened because participants knew the AV would not yield, but decided to speed up rather than let the AV pass: \textit{``I knew the vehicle would not stop earlier than I usually would, I risked it to quickly cross before the car reached''} - P8.

\subsection{\revised{eHMI Impact on the Crossing Experience}} 
\begin{figure*}
    \centering
    \includegraphics[width=\linewidth]{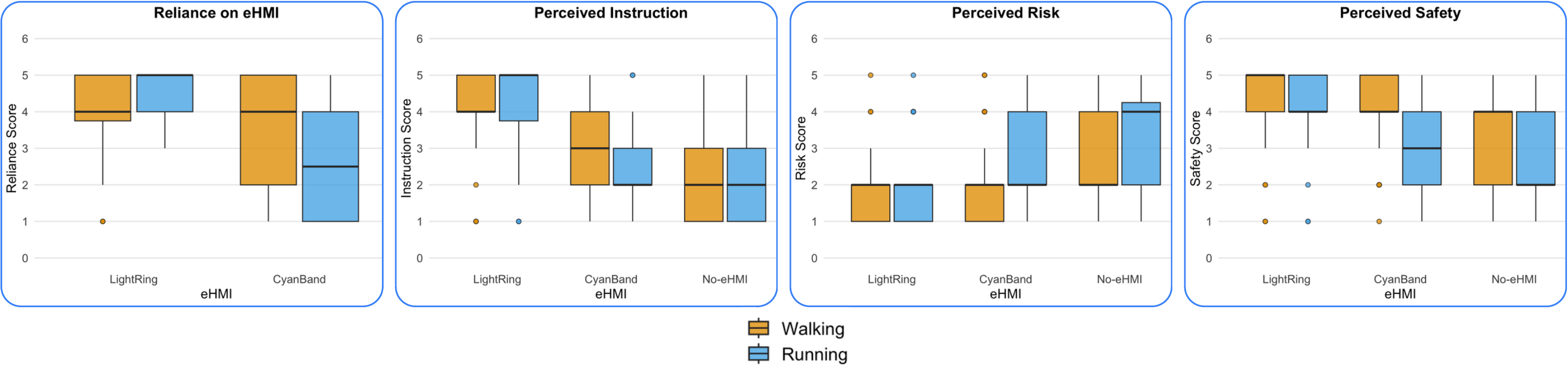}
    \caption{Boxplots showing the distribution of scores for each eHMI across walking (orange) and running (blue) conditions, for the Reliance on the eHMI, Perceived Instruction, Perceived Risk and Perceived Safety subscales. Data for No-eHMI were omitted for the Reliance on the eHMI metric. LightRing was consistently the best-performing, CyanBand performed better for walkers than runners, and No-eHMI underperformed in all metrics.}
    \label{fig:post-crossing-1}
\end{figure*}

\begin{figure*}
    \centering
    \includegraphics[width=\linewidth]{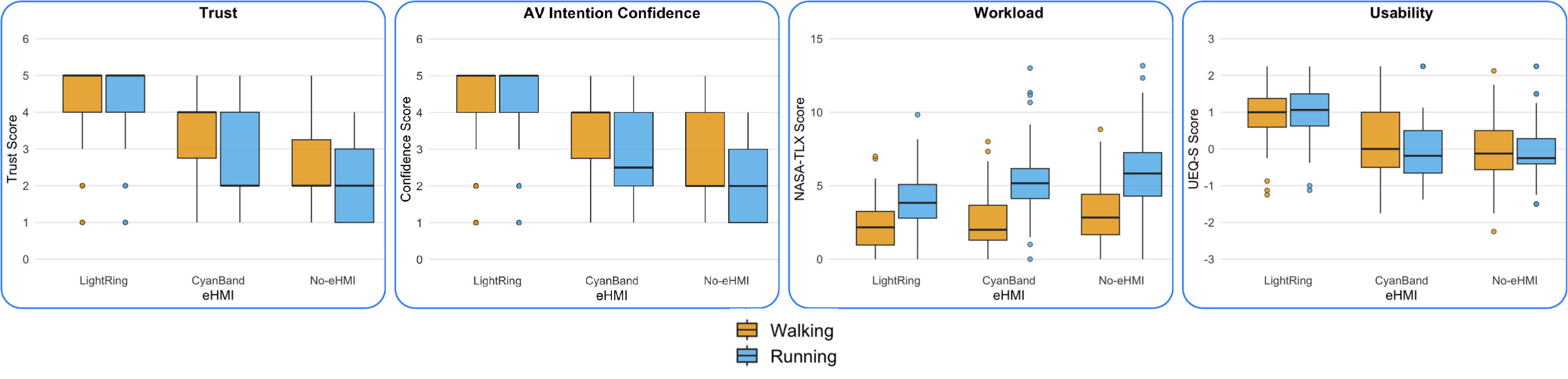}
    \caption{Boxplots showing the distribution of scores for each eHMI across walking (orange) and running (blue) conditions, for the Trust, Confidence in AV Intentions, Crossing Workload and Usability subscales. LightRing was consistently the best-performing, CyanBand performed better for walkers than runners, and No-eHMI underperformed in all metrics.}
    \label{fig:post-crossing-2}
\end{figure*}

This section shows how walker and runner movement behaviours influenced their perceptions and use of each eHMI, as indicated in the post-crossing questionnaire. The data were not normally distributed. Therefore, we conducted the same analysis as that for \textit{Crossing Speed} and \textit{Head Rotations} for each questionnaire subscale. Mean values are in \autoref{tab:post-crossing-mean}. \autoref{fig:post-crossing-1} and \autoref{fig:post-crossing-2} present boxplots illustrating each eHMI's questionnaire scores per activity.


\begin{figure*}
    \centering
    \includegraphics[width=\linewidth]{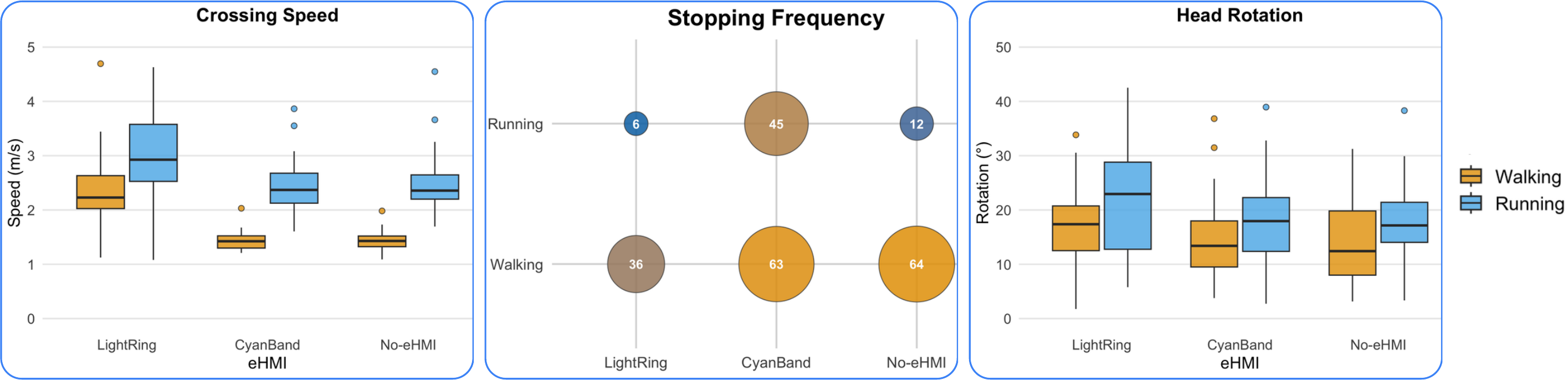}
    \caption{Plots showing pedestrian behaviours when using each eHMI per walking and running conditions. Left: Boxplot of Crossing Speed; Middle: Frequency Plot of Stopping Frequency; Right: Boxplot of Head Rotations. In the Crossing Speed and Head Rotation plots, orange boxes represent walking and blue ones represent running. }
    \label{fig:behaviour}
\end{figure*}

\paragraph{\textbf{Reliance on the eHMI}} Data for the \textit{No-eHMI} condition were excluded because there was no display. There were no effects of \textit{activity} (F(1, 161) = 1.02, $P=.3$) and \textit{AV-behaviour} (F(1, 161) = 0.58, $P=.4$). However, there was a significant main effect of \textit{eHMI} (F(1, 161) = 45.25, $P< .001$; $\eta^{2}$ = 0.22), indicating that participants relied more on \textit{LightRing} than \textit{CyanBand}, \revised{likely because they had faster interactions with LightRing and less time to consider AV driving behaviours}.

\paragraph{\underline{Interactions}} There were significant interactions between \textit{activity and eHMI} (F(1, 161) = 7.25, $P=.008$; $\eta^{2}$ = 0.04), and between \textit{eHMI and AV-behaviour} (F(1, 161) = 5.06, $P=.03$; $\eta^{2}$ = 0.03). \textit{Post hoc} comparisons of the interaction between \textit{activity and eHMI} showed that using \textit{LightRing while running} led to the highest reliance on the eHMI across all eHMI-activity combinations ($p < .05$ for all). Participants also relied significantly more on \textit{LightRing while walking} than on \textit{CyanBand while running} ($p < .0001$). Therefore, runners are more likely to rely on the eHMI than walkers, but the display must meet their specific needs. Comparisons of the interaction between \textit{eHMI and AV-behaviour} showed that participants relied significantly more on the eHMI when using both \textit{LightRing when the AV yielded} and \textit{LightRing when the AV did not yield} than \textit{CyanBand across all AV behaviours} ($p<.005$ for all).

\paragraph{\textbf{Perceived Instruction from the AV}} 
There were significant main effects of all three factors: \textit{activity} (F(1, 253) = 5.71, $P=.02$; $\eta^{2}$ = 0.02), showing that \textit{walkers} felt more instructed than \textit{runners}; \textit{AV-behaviour} (F(1, 253) = 12.99, $P< .001$; $\eta^{2}$ = 0.05), with participants feeling more instructed when the \textit{AV did not yield}, and \textit{eHMI} (F(2, 253) = 76.65, $P< .001$; $\eta^{2}$ = 0.38); with \textit{post hoc} tests showing that participants felt significantly more instructed when using \textit{LightRing} than the others ($p<.0001$ for both), and with \textit{CyanBand} than \textit{No-eHMI} ($p=.002$). 

\paragraph{\underline{Interactions}}  There was a significant interaction between \textit{activity and eHMI} (F(2, 253) = 5.34, $P=.005$; $\eta^{2}$ = 0.04). \textit{Post hoc} tests showed that participants felt more instructed with \textit{LightRing across both activities} than all other eHMI-activity combinations ($p<.005$ for all), and with \textit{CyanBand while walking} than \textit{No-eHMI across both activities} ($p<.005$ for both); this was not observed for \textit{CyanBand while running},  likely due to runners relying less on CyanBand than walkers.

\paragraph{\textbf{Trust}} There were significant main effects of all three factors: \textit{activity} (F(1, 254) = 12.28, $P< .001$; $\eta^{2}$ = 0.05), showing that \textit{walkers} trusted the AV more than \textit{runners}; \textit{AV-behaviour} (F(1, 253) = 5.63, $P=.02$; $\eta^{2}$ = 0.02), with higher trust when the \textit{AV yielded} and its speed changes were clearer; \textit{eHMI} (F(2, 254) = 80.56, $P< .001$; $\eta^{2}$ = 0.39), with \textit{post hoc} tests showing that trust was significantly higher with \textit{LightRing} than the others ($p < .0001$ for both), and \textit{CyanBand} than \textit{No-eHMI} ($p < .0001$). This is likely due to having an explicit display confirming implicit driving behaviours.

\paragraph{\underline{Interactions}}  There was a significant interaction between \textit{activity and eHMI} (F(2, 254) = 6.12, $P=.002$; $\eta^{2}$ = 0.05). \textit{Post hoc} tests showed that trust was significantly higher with \textit{LightRing in either activity} than all other conditions ($p < .05$ for all). Participants also reported higher trust using \textit{CyanBand while walking} than \textit{while running} ($p = .002$) and \textit{No-eHMI across both activities} ($p < .0001$ for both).

\paragraph{\textbf{Confidence in AV Intentions}} There was a significant main effect of \textit{activity} (F(1, 254) = 17.25, $P< .001$; $\eta^{2}$ = 0.06); \textit{walkers} were more confident in the AV's intentions than \textit{runners}, potentially because they had more time to decide. However, there was no effect of \textit{AV-behaviour} (F(1, 253) = 1.49, $P=.2$). There was a significant main effect of \textit{eHMI} (F(2, 254) = 81.55, $P< .001$; $\eta^{2}$ = 0.39), and \textit{post hoc} tests showed that confidence was significantly higher with \textit{LightRing} than the others ($p<.0001$ for both), and \textit{CyanBand} than \textit{No-eHMI }($p<.0001$). Therefore, implicit cues were insufficient. 

\paragraph{\underline{Interactions}} There was a significant interaction between \textit{activity and eHMI} (F(2, 254) = 4.61, $P=.01$; $\eta^{2}$ = 0.04). \textit{Post hoc} tests showed that \textit{LightRing while walking or running} outperformed all other eHMI-activity combinations ($p < .05$ for all). \textit{CyanBand while walking} resulted in significantly higher confidence than using it \textit{while running} ($p = .003$) and \textit{No-eHMI across both activities} ($p < .001$ for both). Therefore, CyanBand performed better for walkers than runners.

\paragraph{\textbf{Perceived Risk}} There was a significant main effect of \textit{activity} (F(1, 253) = 16.54, $P< .001$; $\eta^{2}$ = 0.06); \textit{runners} felt they made riskier crossing decisions than \textit{walkers}. There was no effect of \textit{AV-behaviour} (F(1, 253) = 0.05, $P=.8$). There was a significant main effect of \textit{eHMI} (F(2, 253) = 36.32, $P< .001$; $\eta^{2}$ = 0.22), and \textit{post hoc} comparisons showed that decisions were perceived as significantly less risky with \textit{LightRing} than the others ($p<.0005$ for both) and with \textit{CyanBand} than \textit{No-eHMI} ($p<.0001$). Therefore, pedestrians need explicit displays for informed crossing decisions.

\paragraph{\underline{Interactions}} There was a significant interaction between \textit{activity and eHMI} (F(2, 253) = 4.47, $P=.01$; $\eta^{2}$ = 0.03). \textit{Post hoc} comparisons showed \textit{No-eHMI while running} led to significantly riskier decisions than all other eHMI-activity combinations ($p < .05$ for all). \textit{LightRing at either activity} outperformed \textit{CyanBand while running} and \textit{No-eHMI at both activities} ($p < .005$ for all). There were no significant differences for LightRing between walking and running; it was inclusive. In contrast, using \textit{CyanBand while running} resulted in riskier decisions than \textit{while walking} ($p = .009$).

\paragraph{\textbf{Perceived Safety}} There was a significant main effect of \textit{activity} (F(1, 253) = 25.21, $P< .001$; $\eta^{2}$ = 0.09), showing that \textit{walkers} felt safer than \textit{runners}. There was no effect of \textit{AV-behaviour} (F(1, 253) = 2.72, $P=.1$). However, there was a significant main effect of \textit{eHMI} (F(2, 253) = 46.95, $P< .001$; $\eta^{2}$ = 0.27); \textit{post hoc} tests showed perceived safety was significantly higher with \textit{LightRing} than the others ($p<.0001$ for both), and with \textit{CyanBand} than \textit{No-eHMI} ($p=.009$).

\paragraph{\underline{Interactions}}  There were significant interactions between \textit{activity and eHMI} (F(2, 253) = 5.9, $P=.003$; $\eta^{2}$ = 0.04), \textit{activity and AV-behaviour} (F(1, 253) = 4.39, $P=.04$; $\eta^{2}$ = 0.02), and \textit{eHMI and AV-behaviour} (F(2, 253) = 3.77, $P=.02$; $\eta^{2}$ = 0.03). \textit{Post hoc} comparisons of the \textit{activity and eHMI} interaction revealed that participants felt significantly safer using \textit{LightRing in both activities} and \textit{CyanBand while walking} than \textit{CyanBand while running} and \textit{No-eHMI in both activities} ($p < .05$ for all). They also felt safer using \textit{No-eHMI while walking} than \textit{while running }($p = .02$). 

\textit{Post hoc} comparisons of the \textit{activity and AV-behaviour} interaction showed that both \textit{walking when the AV yielded} and \textit{walking when the AV did not yield} were perceived as significantly safer than \textit{running under all AV behaviours} ($p < .05$ for all). Comparisons of the interaction between \textit{eHMI and AV-behaviour} showed that participants felt significantly safer using \textit{LightRing in both yielding conditions} than all other eHMI-yielding behaviour combinations ($p<.001$ for all). Using \textit{CyanBand when the AV yielded} was also perceived to be significantly safer than \textit{No-eHMI when a non-yielding AV} ($p=.02$). Therefore, participants felt safer interpreting CyanBand only when the AV yielded, but LightRing in all cases.

\paragraph{\textbf{\revised{Crossing Workload (NASA-TLX)}}} There was a significant main effect of \textit{activity} (F(1, 253) = 150.74, $P< .001$; $\eta^{2}$ = 0.37), with \textit{runners} experiencing a higher workload than \textit{walkers}. However, there was no effect of \textit{AV-behaviour} (F(1, 253) = 0.98, $P=.3$). There was a significant main effect of \textit{eHMI} (F(2, 253) = 17.39, $P< .001$; $\eta^{2}$ = 0.12), and \textit{post hoc} comparisons showed that workload was significantly lower with \textit{LightRing} than the others ($p<.005$ for both).

\paragraph{\underline{Interactions}} There was a significant interaction between \textit{activity and eHMI} (F(2, 253) = 5.34, $P=.005$; $\eta^{2}$ = 0.04). \textit{Post hoc} tests showed using any eHMI \textit{while running} generated a significantly higher workload than using it \textit{while walking} ($p < .001$ for all). There were no significant differences between eHMIs during walking. However, while running, \textit{LightRing} outperformed the other eHMIs ($p < .001$ for both). \textit{CyanBand and No-eHMI while running} also caused a higher workload than all walking conditions ($p < .0001$ for all), so LightRing was particularly effective for runners.

\paragraph{\textbf{Usability (UEQ-S)}} There was no effect of \textit{activity} (F(1, 253) = 0.74, $P=.4$). However, there was a significant main effect of \textit{AV-behaviour} (F(1, 253) = 3.91, $P=.05$; $\eta^{2}$ = 0.02), with the interaction being more usable when the \textit{AV yielded}. This is likely because the AV was slower and its intentions were more apparent through its driving behaviour. There was also a significant main effect of \textit{eHMI} (F(2, 253) = 62.95, $P< .001$; $\eta^{2}$ = 0.33), and \textit{post hoc} comparisons showed that \textit{LightRing} was significantly more usable than the others ($p<.0001$ for both). There were no interactions.

\subsection{\revised{Qualitative Insights on the Crossing Experience}}

\begin{figure*}
    \centering
    \includegraphics[width=\linewidth]{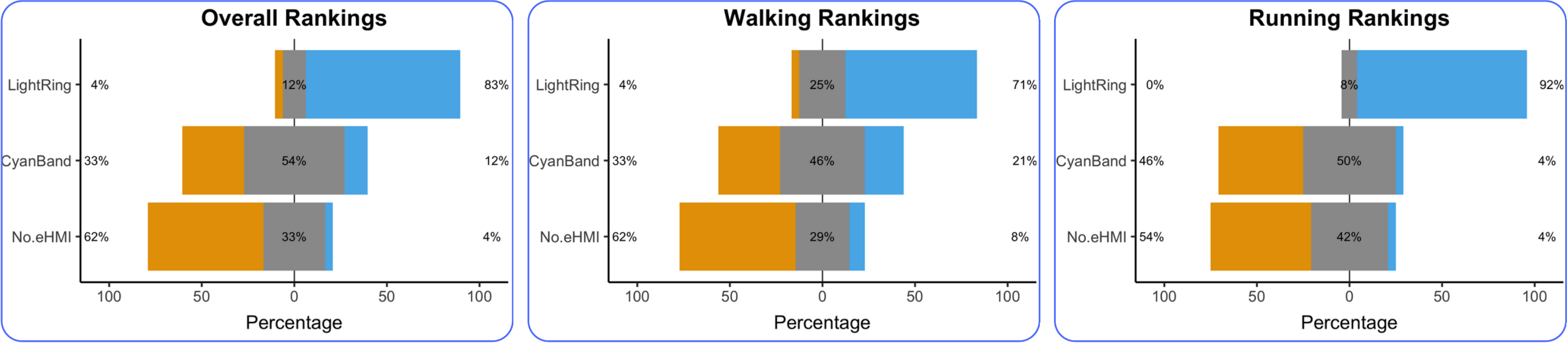}
    \caption{Participant rankings of the eHMIs. Left: Overall rankings across walking and running; Middle: Rankings for the walking condition; Right: Rankings for the running conditions. Orange bars toward the left represent lower rankings, Grey bars in the middle represent neutral rankings, and blue bars on the right represent higher rankings.}
    \label{fig:rankings}
\end{figure*}

\autoref{fig:rankings} shows participant eHMI rankings overall and then separately per activity. \textit{LightRing} was ranked as the best and \textit{No-eHMI} as the worst in all cases. \textit{LightRing} performed particularly well for runners; no participants ranked it as the worst condition. \textit{CyanBand} was received differently: \textit{Walkers} were more positive, with 21\% ranking it as the best condition. However, only 4\% of runners ranked \textit{CyanBand} as the best, while 46\% ranked it as the worst.

\subsubsection{\textbf{Themes}} An inductive, data-driven thematic analysis \cite{Braun2006UsingPsychology} was conducted on the qualitative (free-text) feedback provided by participants after each trial and at the end of the study. The data were imported into NVivo\footnote{NVivo qualitative analysis software: \url{lumivero.com/products/nvivo/} (Accessed: 31/08/2025)} for coding. One author identified 24 unique codes. Two authors then collaboratively sorted these into five overarching themes based on their similarities. The process was iterative, with disagreements discussed and resolved as needed. Codes were remapped when necessary, and themes containing overlapping codes were reassessed and combined where appropriate.

\paragraph{\textbf{Theme 1: Factors differentiating runners from walkers}} We identified three differentiating factors:
\begin{itemize}
    \item \textit{\underline{Purpose of crossing:}} Pedestrians typically run to improve physical health: \textit{``I run for exercise; it is important that I keep running and don’t stop''} - P4. In contrast, walking is generally a commuting or leisurely activity: \textit{``I walk to work or take a stroll around the city. I don’t mind stopping to let cars pass; there’s no rush''} - P9. Therefore, the crossing purpose differs; runners primarily cross to maintain exercise, preferring not to stop for vehicles, while walkers cross to reach a destination at a comfortable pace.
    \item \textit{\underline{Available time:}} Runners have less time to make crossing decisions than walkers: \textit{``Time makes all the difference. Everything becomes harder when running. I must quickly check the display on the car while trying to process its speed without affecting my running pace''} - P7; \textit{``You have more time when walking, so you can check the eHMI and vehicle speed more easily''} - P13. Therefore, available time impacts eHMI usability;
    \item \textit{\underline{Workload:}} Running consumes more physical and mental resources than walking, affecting crossing decisions: \textit{``I'm more tired when I run. eHMIs are useful, but they need to be low-effort to process''} - P12. Therefore, eHMIs should use readily distinguishable signals to best support runners.
\end{itemize}

\paragraph{\textbf{Theme 2: Colour-changing eHMI signals}} LightRing's colour changes were useful for quickly understanding AV intentions:  \textit{``Red and green lights made it easy to understand what it was going to do, and it was easy to recognise that at a glance''} - P1. These were particularly useful while running, where there is less time to cross: \textit{``I preferred the red/green display in both movement tasks, but especially when running'' } - P6. However, there is a risk of over-reliance on the eHMI: \textit{``Red and green require a lot of trust in the runner. I’d worry it might be too risky, but it was best for me''} - P15. Therefore, red and green were highly distinguishable, but took significant precedence over vehicle behaviour.

\paragraph{\textbf{Theme 3: Animation for eHMIs}} 
LightRing used flashing/pulsing animations in addition to colour changes, but participants relied primarily on the colours: \textit{``I liked the changes in colour. Didn't pay attention to the flashing''} - P11. Some felt that LightRing's animations added ambiguity: \textit{``I prefer solid colours; animations added a layer of complexity''} - P12; \textit{``[pulsing] green confused me. It made me feel more rushed as if the car will move imminently''} - P24. This further shows that colours outperform animations for distinguishable signals. CyanBand's animations were more useful for walkers than runners: \textit{``When walking, I had more time to think, the cyan lights indicating movement helped me understand what the car was doing''} - P9; \textit{``When running, cyan lights were difficult to understand because directionality isn’t clear until you’ve watched it for a while''} - P1. Therefore, CyanBand required continuous monitoring, reducing its inclusivity. 

\paragraph{\textbf{Theme 4: eHMI placement}} CyanBand was placed only on the front; participants struggled to easily spot the eHMI: \textit{``Cyan lights were difficult to spot, particularly when running. They are relatively thin lines which only appear on the front of the car''} - P8; \textit{``CyanBand's display location was suboptimal aka too small''} - P24. Some explained that this depended on the AV's yielding behaviour: \textit{``If the vehicle stopped, then the user has a chance to understand the small interface in front''} - P2. Therefore, participants preferred broader eHMI visibility to easily find and process signals, as in LightRing.

\paragraph{\textbf{Theme 5: Extending eHMIs to personal devices}} Runners already use devices such as smartwatches or earphones, which could be connected to eHMIs to extend their signals: \textit{``Audio prompt from earphones would be good. Haptics on a smartwatch could also be useful''} - P10;  \textit{``Smart devices help runners with notifications and some vibrations''} - P19. Therefore, pedestrians are open to non-visual cues from their personal devices, in addition to visual eHMIs.

\section{Limitations and Future Work}

\revised{We used an AR simulator rather than a real vehicle to eliminate physical risk when participants experienced non-yielding AV scenarios. We ensured the validity of our setup using an established AR simulator that was shown to elicit realistic road-user behaviours and high levels of presence in prior studies \cite{Al-Taie2025AroundCultures, Al-Taie2025EvARythingInterfaces}. Nevertheless, future work is encouraged to investigate new ways of safely exposing participants to non-yielding AV behaviours using real vehicles.} \revised{Our findings were derived from a sample recruited from the local area. While we ensured that this sample comprised balanced genders, a wide age range, and a mix of regular and casual runners, future work should broaden this demographic to older adults, children, and runners with accessibility needs to further support inclusive eHMI design \cite{Hollander2021AReview, Abe2025IRunners}.} \revised{Since our study took place in the UK using UK-specific road layouts, cross-cultural replications are needed to determine whether runners in other regions require different eHMI design considerations based on local traffic norms \cite{Al-Taie2025AroundCultures}.} 

The AV in our study was a Citroën C3, a small city car. Future work should test the same eHMIs on different vehicle sizes, such as SUVs, trucks, or buses, as pedestrians may respond differently around them \cite{Lau2022OnePerspective}. As a starting point and for comparison with prior work \cite{Dey2020TamingInterfaces}, we evaluated the eHMIs on an SAE Level 5 AV without passengers \cite{SocietyforAutomotiveEngineers2021SAEVehicles}. Future studies should investigate AVs of varying SAE automation levels, including those with a visible passenger in the driving seat, since pedestrians could be unsure whether to rely on the eHMI or the human occupant \cite{Dey2020TamingInterfaces}. We focused on one-to-one AV–pedestrian interactions as an initial step. However, eHMIs must remain effective in one-to-many or many-to-many interactions for real-world deployment \cite{Tran2023ScopingInteraction}. Both the LightRing and CyanBand have been tested in scalable settings \cite{Al-Taie2025EvARythingInterfaces, Dey2021TowardsPedestrian}, but not with runners. Future research should explore AV–runner interaction in multi-vehicle settings, where runners must divide attention between multiple AVs despite having limited time to cross \cite{Tran2022DesigningInteraction}.
\section{Discussion and Design Guidelines}

\revised{We addressed our RQs by demonstrating how the embodied differences between walkers and runners influenced their crossing behaviours and eHMI use during AV interactions. Our findings demonstrate that AV-pedestrian interaction is fundamentally shaped by embodied movement patterns, motivation, and situational constraints. Walkers and runners utilise time, attention, and physical capability differently during crossing decisions, resulting in distinct patterns of interface reliance, risk tolerance, and trust calibration. Building on this empirical evidence, we propose the following design guidelines (DGs) and use them to connect our results with theories on embodied cognition, situational impairment, and human–machine communication, thereby supporting more inclusive eHMI design}

\paragraph{DG1: Researchers must understand the motivation behind the user's physical activity} The nature of AV interactions differed between walkers and runners. Walkers interacted with the AV in a manner similar to stationary participants \cite{Dey2021CommunicatingBehavior}: they were more likely to slow down or stop at the crossing, giving them sufficient time to interpret eHMI signals and verify them against the AV’s driving behaviours \cite{Bazilinskyy2021HowParticipants}. This resulted in higher trust and confidence in the AV’s intentions. In contrast, runners had faster, more mobile interactions. They maintained pace and attempted to interpret eHMI signals while moving, even when this led to risky manoeuvres. These findings align with prior work suggesting that runners prefer to avoid crossings so they can maintain continuous movement \cite{Deelen2019AttractiveEnvironment., Ettema2016RunnableFrequency}, but this is not always possible in urban areas \cite{Urmson2008AutonomousChallenge}. Our results also mirror real-world observations of driver–walker interactions, where walkers frequently stop before crossing \cite{Dey2017TheVehicles}. \revised{Therefore, ARcade elicited realistic behaviours from participants.} 

\revised{Our study adds important context to these findings; the observed differences stem partly from the fundamental purpose of crossing being different between walking and running. Participants described walking as a commuting or leisurely activity, whereas running was framed as an exercise activity tied to performance goals such as reaching a target distance or time. This has broader implications for HCI research that examines device use in active settings. Prior work has given limited attention to why people engage in these activities or the goals that shape their behaviour \cite{Matviienko2022BabyBike, Winters2012Safe, HegnaBerge2023SupportOutlook}. For example, cycling is often studied as a single category even though it can serve both commuting and sports purposes \cite{Al-Taie2022TourBehaviour}. Our results demonstrate that these underlying goals meaningfully influence how people use interfaces in practice. We recommend that studies evaluating interfaces in active settings explicitly collect data (e.g., through interviews) about the use cases in which participants would normally perform the activities and use the tested systems. Researchers should also incorporate real-world contextual elements into experimental setups to help participants draw on the lived experience of practising these activities, such as ARcade’s ability to overlay content onto a real environment.}

\paragraph{DG2: Physical constraints shape interface use between activities} \revised{The differences between walking and running movements also shaped the types of AV interactions. Sports science research shows that runners require substantially more effort than walkers to slow down and accelerate again \cite{Alexander1996WalkingRunning, Jin2024LowerTransitions}, and our participants reported similar constraints. To maintain their running rhythm, participants performed larger head rotations while moving, rather than slowing down and altering their whole-body movement \cite{Yao2022RecognizingSpeeds}. Previous research demonstrated that running interfaces must minimise interference with movement \cite{Seuter2017RunningTechnology}, which is why runners prefer constrained form factors such as small smartwatch screens \cite{Hamdan2017RunTap:Runners}. However, we showed that this applies to safety-critical situations, not just fitness tracking.} 

\revised{In contrast, walkers were more accepting of adapting their whole-body movement to interpret AV intentions. Our findings with CyanBand showed that slowing down or altering gait while walking does not significantly hinder the interface's performance \cite{Bergstrom-Lehtovirta2011TheInterface}. This observation is not limited to road safety; broader research showed that walkers can still complete complex tasks, such as texting, reading, or target selection, on smartphones or AR headsets and continue to perceive these interfaces as usable despite temporary reductions in speed \cite{Bergstrom-Lehtovirta2011TheInterface, Chang2024ExperienceScenarios, Licence2015GaitObstacles}. Recent walker-centric interfaces were even intentionally designed to induce changes in movement, such as anchoring AR content to the user’s arm \cite{Rasch2025ARWalking} or supporting input through body gestures like kicking \cite{Tsai2024GaitWalking}. These results show that it is insufficient to classify interface users only based on their movement speed \cite{Hollander2021AReview}; their distinct movement patterns and physical constraints are crucial factors that shape interface use. Therefore, it is essential to begin exploring how running movement influences the usability of such displays, particularly as emerging technologies, such as extended reality, take on sports-oriented form factors \cite{vonSawitzky2022HazardStudy}, opening new doors for runner-based interactions.} 

\paragraph{DG3: Study designs must facilitate exposure to all interface states} Real-world observations of human driver-walker encounters found that walkers could safely cross without explicit interaction \cite{Moore2019TheVehicles, Lee2021RoadVehicles}. These works concluded that implicit cues may be sufficient for AV interactions. However, the No-eHMI condition consistently underperformed for both walkers and runners in our study. It imposed a higher workload and risk, and lower usability, and confidence in AV intentions than any eHMI. In contrast, the inclusion of an eHMI also improved pedestrian crossing behaviours, as runners and walkers were more comfortable maintaining their pace with LightRing. These results align with prior work investigating other road user types \cite{Matviienko2022E-ScootAR:Reality, Rettenmaier2020AfterScenarios}, reinforcing the idea that it is essential to design inclusive eHMIs \cite{Hollander2021AReview} and validating our work in this paper, which considers runners for the first time.

\revised{However, safe crossing decisions rely on a combination of explicit and implicit AV signals \cite{Markkula2020DefiningTraffic}. Therefore, eHMIs must work across all AV behaviours. We found that the crossing experience was hindered by a non-yielding AV; participants were more susceptible to collisions and felt instructed to stop rather than negotiate their right of way. With CyanBand, participants felt safer only when the AV yielded; it is insufficient to expose participants only to yielding AVs when testing eHMIs. This makes it challenging to draw valid conclusions from studies using methods such as Ghost Driver \cite{Rothenbucher2015GhostDriver}, which requires a real vehicle and necessarily omits hazardous non-yielding behaviours \cite{Al-Taie2024LightInterfaces}. 
Situational impairment theory proposes that movement and environmental conditions can temporarily constrain a user's ability to engage with an interface \cite{Marentakis2024SituationalInterfaces}. We extend this to eHMIs, and show that the AV's driving behaviour could impair AV-pedestrian interactions \cite{Sarsenbayeva2017ChallengesDevices}. Therefore, future research should ensure that participants safely experience all AV behaviours to assess usability in all situations. This could be done through simulators, such as ARcade.}

\paragraph{DG4: Movement influences cross-checking interface signals with motion cues} \revised{Previous research on safety-critical human-machine communication showed that users must validate interface signals against real-world cues. For example, \citeauthor{Sauppe2015TheSettings}~\cite{Sauppe2015TheSettings} found that factory workers in hazardous environments only trusted a robot’s interface when its signals aligned with robotic motion. Our findings extend this by showing how a human user's movement behaviour shapes their ability to cross-check signals. Walkers' slower movements made it straightforward to verify eHMI signals with the AV's motion. However, runners had little opportunity to cross-check. Consequently, they were prone to unsafe crossings when relying solely on the eHMI, with LightRing, or on implicit cues alone, with No-eHMI.}

\revised{Research on runners' use of wearable fitness trackers showed that they rarely relied on the display alone, and often cross-checked device feedback with embodied cues, such as breathing rate \cite{Seshadri2019WearableAthlete, Clermont2020RunnersInjury}. Therefore, cross-checking can be supported without conflicting running movement. However, it becomes challenging when the information at stake, such as AV intention, cannot be inferred from bodily feedback. Nevertheless, our qualitative data suggest that AV intentions could be delivered locally through common wearable devices used for running. Smartwatches or earphones could optionally connect to eHMIs and provide localised cues, helping runners map external signals onto their embodied readiness to sprint or slow down. This would also reduce the need for large head rotations or other disruptive movements \cite{Al-Taie2024BikeInterfaces, Seuter2017RunningTechnology}. Similar approaches have been successfully explored with cyclists through bike computer integration \cite{Al-Taie2024BikeInterfaces}. Our results suggest that an equivalent strategy is viable for runners.}

\paragraph{DG5: Mobile interfaces must be tested with moving participants} CyanBand was not inclusive of pedestrian behaviours; walkers were more receptive, but runners did not rely on the eHMI. This was due to uni-colour directional animations being challenging to interpret while in motion, and the eHMI's placement solely on the vehicle's front. Prior work with cyclists also showed that they prefer more distinguishable signals and broader eHMI visibility \cite{Al-Taie2024LightInterfaces}. Therefore, our results align runners more closely with cyclists than with walkers. Both runners and cyclists interact with AVs while moving \cite{Al-Taie2023KeepTraffic}. However, cyclists do this out of necessity; they share the road with vehicles and must interact with them in scenarios such as lane merging \cite{Hou2020AutonomousPromise}, whereas runners prefer to maintain pace, despite being able to stop.  

CyanBand still showed limitations with walkers, as they slowed down considerably to interpret its signals, even more than the No-eHMI condition. These findings were not captured in prior work, which tested CyanBand with stationary participants \cite{Dey2020TamingInterfaces, Dey2020ColorPedestrians}. \revised{Therefore, our results emphasise the importance of having moving participants and capturing their movement data to obtain a holistic understanding of interface usability. Multiple resource theory \cite{Wickens2008MultipleWorkload} explains that physical activity limits perceptual and attentional resources, reducing situational awareness and impairing the performance of concurrent tasks such as interpreting eHMIs. Prior work built on this idea to demonstrate that walkers maintain situational awareness by dividing attention between interfaces and the surrounding environment, intentionally focusing on display content for a few seconds at a time \cite{Oulasvirta2005InteractionBursts}. For eHMIs, we showed that CyanBand required continuous monitoring to determine animation directions, and led to walkers slowing down significantly, possibly due to reduced awareness of their environment. Our results with CyanBand confirm that it is insufficient to test eHMIs only with walkers. Previous research classified walkers as the slowest road users \cite{Hollander2021AReview}, and we showed that eHMIs designed for slower-paced road users do not necessarily generalise to faster-paced ones. However, the success of LightRing suggests that displays developed for faster road users may generalise more effectively to slower ones.}

\paragraph{DG6: High distinguishability can cause over-reliance on interfaces} LightRing was the most positively received eHMI between walkers and runners, aligning with prior work showing that colour changes provide highly distinguishable signals for rapid interpretation, even for faster road users such as drivers \cite{Dey2020TamingInterfaces, Hollander2021AReview, Matviienko2022E-ScootAR:Reality, Al-Taie2025AroundCultures}. Therefore, LightRing is a potentially inclusive solution. \revised{However, previous research did not examine how LightRing is used alongside the AV's implicit cues \cite{Al-Taie2024LightInterfaces, Al-Taie2025AroundCultures, Al-Taie2025EvARythingInterfaces}, and we identified limitations in this area. With walkers, LightRing sped up the alignment of explicit and implicit signals, so they still considered the AV's driving behaviours. However, runners reported an over-reliance on the eHMI and a lack of consideration for implicit cues. Here, rapid interpretation was favoured over informed decisions so runners could maintain their pace, even with non-yielding AVs. This led to reckless and collision-prone behaviours.} 

\revised{These findings raise important questions about balancing highly distinguishable interface states with informed decision-making in safety-critical contexts. Prior work explored this in stationary contexts, such as digital nudging in online forms \cite{Schneider2018DigitalDesign, Hettler2025UnderstandingDesign} or human user autonomy in AI collaboration \cite{Papantonis2025WhyCollaboration}. These works recommend toning down highly salient colours or adding uncertainty cues in autonomous systems to prompt users to verify what they see. However, crossing the road is fundamentally different: signals must be highly noticeable, unambiguous, and confidence-inducing \cite{Markkula2020DefiningTraffic, Al-Taie2023KeepTraffic, Mahadevan2018CommunicatingInteraction}. With LightRing, one possible mitigation strategy is delaying the display of explicit signals, prompting road users to first process implicit cues. However, designers must avoid long delays that would diminish runners' reliance on the eHMI because runners selectively use interfaces based on their needs, as we found with CyanBand. Therefore, a prolonged delay may cause runners to ignore the interface and cross rapidly to maintain their pace. Future designers would need to balance these trade-offs for effective eHMI design.}

\begin{figure*}
    \centering
    \includegraphics[width=0.8\linewidth]{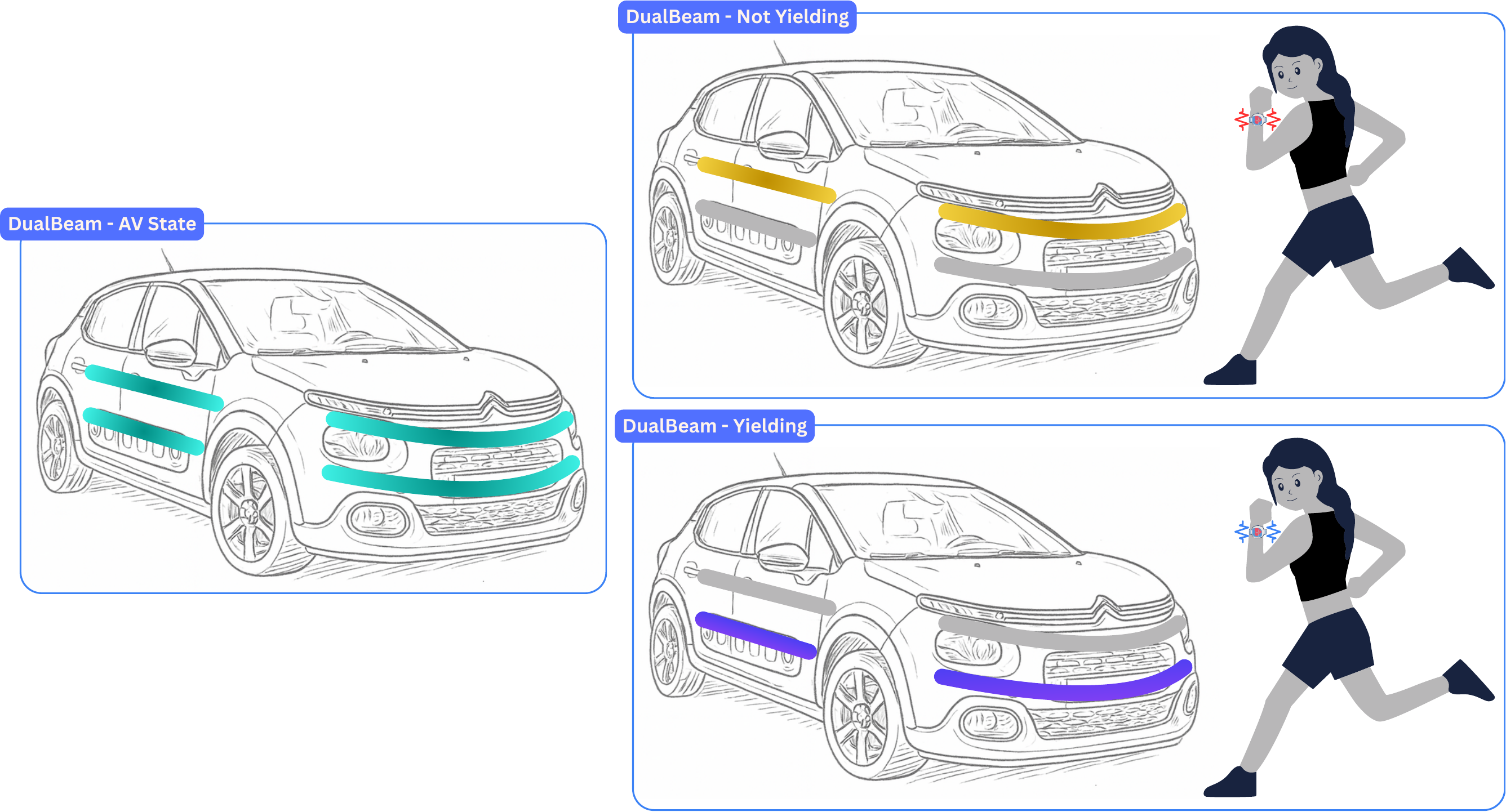}
    \caption{DualBeam: our novel AV interface based on the design guidelines. Left: DualBeam communicating the vehicle's autonomous state with both lightbars on in cyan; Top-right: The top lightbar on in amber and pulsing smartwatch vibration to communicate not-yielding; Bottom-right: The bottom lightbar on in purple and continuous smartwatch vibration to communicate yielding.}
    \label{fig:dualbeam}
\end{figure*}

\paragraph{DG7: Embodied behaviours shape perceived autonomy in human-machine communication} Participants felt most instructed with LightRing. Previous research suggested this could be due to the use of red and green, which are associated with traffic lights \cite{Dey2020ColorPedestrians}. However, perceived instruction was also high with CyanBand, meaning the introduction of explicit signals heightens this effect. Runners did not rely on CyanBand, which could explain why the effect was stronger with LightRing. \revised{We argue that perceived instruction is shaped by the interaction speed and the user's motivation to cross, rather than solely by the eHMI's colours. Walkers were slower than runners and slowed further to assess the perceived risk of crossing. This diminished any extra time for safe crossing with a non-yielding AV and removed the opportunity for right-of-way negotiation. In contrast, runners had greater motivation to maintain pace. They could still cross in time with a non-yielding AV, and even proceeded when LightRing displayed red lights to indicate non-yielding. This shows that the road user's movement speed potentially determines their perceived instruction from AVs.} 

\revised{Drawing on embodied cognition theory \cite{Kirsh2013EmbodiedDesign}, another explanation could be how runners use their bodily cues to assess risk. Runners heavily rely on the body's felt sense to infer subtle cues about fatigue and emotional state that wearable devices do not capture \cite{Tholander2025TheRunning}. Applying this to our crossing context, runners may have also drawn on these somatic cues to assess bodily readiness in real-time, deciding how risky it was to sprint across the road. Therefore, runner crossing decisions may be partly shaped by kinaesthetic, embodied judgements grounded in how they feel in the moment \cite{Tholander2025TheRunning}.} Future work should still compare the perceived instruction of alternative contrasting colours with red/green in LightRing. This was done for uni-colour eHMIs, like CyanBand \cite{Dey2020ColorPedestrians}, but we showed that contrasting colours are more effective if carefully designed to help road users make safe, well-informed crossing decisions.

\section{DualBeam: Novel and Inclusive AV interface based on the design guidelines}

To demonstrate how our design guidelines can be applied in practice, we developed DualBeam; a novel AV interface for road users with \revised{distinct motivations, movement behaviours and interaction speeds.} We also incorporated findings from prior work on other road user types, such as cyclists and drivers \cite{Hollander2021AReview}, to help ensure DualBeam's inclusivity beyond walkers and runners. \revised{\textit{DG1} showed that the motivations of the road user shape their interactions with the AV, and \textit{DG5} showed that concepts developed for faster interactions may generalise more effectively than those developed for slower ones. Therefore, we started developing DualBeam with runners in mind, as they are more motivated to cross and interact more quickly than walkers.}  \textit{DG5} explained that eHMIs placed on larger vehicle areas support broader visibility and faster, more informed crossing decisions. Therefore, \textit{DualBeam} is placed around the vehicle, similar to LightRing. This promotes inclusivity by allowing road users to interact with AVs from all directions, particularly useful for those who will consistently share the road with AVs, such as drivers \cite{Hollander2021AReview, Rettenmaier2020AfterScenarios, Al-Taie2023KeepTraffic}. \textit{DG5} also states that animations alone are insufficient for supporting crossing decisions, and our findings showed they could cause confusion when paired with colour changes. Therefore, \textit{DualBeam} conveys intentions through light positions, featuring two rows of light bars: top and bottom. This approach was recommended for cyclists \cite{Al-Taie2024LightInterfaces}. However, it was never tested in practice. It may be effective for pedestrians and even for other road users, since road users can maintain distinct positions at traffic lights while moving at higher speeds \cite{Al-Taie2023KeepTraffic}. Therefore, applying this to eHMIs could mitigate excessive stopping at crossings.

\textit{DG6} demonstrated that eHMI signals must be distinguishable, and  \textit{DG3} showed that this distinguishability should be regardless of AV or road user speed. \textit{DG7} explained that colour changes are promising. However, we recommended investigating contrasting colours beyond red/green to minimise any potential effects of perceived instruction \cite{Dey2020ColorPedestrians} and over-reliance on the eHMI. Therefore, \textit{DualBeam} indicates not-yielding using amber lights on the top bars and yielding using purple lights on the bottom bars. We aligned the bar positions with traffic-light conventions (top for non-yielding, bottom for yielding) to reduce the learning curve and support colourblind users. The amber/purple colour scheme could support a range of road user types, as previous work showed that contrasting colours are effective for communication with VRUs \cite{Hou2020AutonomousPromise} and motorised road users \cite{Dey2020TamingInterfaces, Rettenmaier2020AfterScenarios}. We reused amber because it is traditionally used to warn road users or even to promote negotiation (e.g., through directional indicators) \cite{UK2024TheRules, Al-Taie2023KeepTraffic}, rather than to determine right of way, as with red and green \cite{Dey2020ColorPedestrians, UK2024TheRules}. Purple is new to traffic, potentially mitigating any effects of instruction \cite{Dey2020ColorPedestrians}, and runners could place greater emphasis on AV behaviour while they get accustomed to purple. \textit{DG4} explained that sending eHMI signals to wearables could help runners map these to their own bodily cues and assess their readiness to cross. Accordingly, runners can optionally pair \textit{DualBeam} with their smartwatch to vibrate continuously when the AV yields and pulse when it does not. This can also be customised to personal needs; for example, disabling vibrations with yielding AVs and only including them for non-yielding ones. \textit{DualBeam} can also be integrated with a variety of personal devices to support other road user types and overall inclusivity, such as smartphones or bike computers \cite{Al-Taie2024BikeInterfaces}.
\section{Conclusion}

We conducted the first study examining interactions between AVs and pedestrians with different movement behaviours. Participants used an AR simulator to navigate a virtual crossing while either walking or running. An approaching AV displayed one of three eHMI conditions: LightRing \cite{Al-Taie2024LightInterfaces}, CyanBand \cite{Dey2020ColorPedestrians}, or No-eHMI. \revised{When walking, participants} were more likely to stop at the crossing than runners. This gave them sufficient time to validate eHMI signals with AV driving behaviour, resulting in higher trust and perceived safety. \revised{In contrast, when running, participants} experienced a higher workload overall. They made riskier crossing decisions, reducing confidence in AV intentions and increasing susceptibility to collisions. These differences influenced how participants used eHMIs between the two activities. The No-eHMI condition consistently underperformed across both activities, highlighting the need for explicit AV signals. Colour-changing lights in LightRing were the most effective across both walking and running, providing a promising common solution. However, animated cyan lights from CyanBand were only effective while walking, as they required more time to process. These findings demonstrate that it is crucial to test AV interfaces across a range of road user behaviours to ensure inclusivity. Our results support the large-scale adoption of AVs and help preserve runner safety in future traffic, aligning with the UN's Sustainable Development Goals \cite{UnitedNations2025SustainableGoals} on physical health and wellbeing.

\begin{acks}
Ammar Al-Taie is supported by the KAIST Jang Young Sil Fellowship Program (Excellence Track). The authors acknowledge support from the IITP (Institute of Information \& Communications Technology Planning \& Evaluation)-ITRC (Information Technology Research Center) grant funded by the Korean government (Ministry of Science and ICT) (IITP-2026-RS-2024-00436398), and from the European Research Council (ERC) under the European Union’s Horizon 2020 research and innovation programme (\#835197, ViAjeRo).
\end{acks}

\bibliographystyle{ACM-Reference-Format}
\bibliography{references}

\end{document}